\documentclass[twocolumn,tightenlines,showpacs,prd,floatfix,preprintnumbers,amsmath,amssymb,nofootinbib,superscriptaddress]{revtex4-1}

\usepackage{color}
\usepackage{graphicx, subfigure}
\usepackage{float}
\usepackage{amsmath}
\usepackage{acronym}
\usepackage{multirow}
\usepackage{amsfonts}
\usepackage{ulem}

\usepackage{color}
\newcommand{\av}{\textcolor{black}}
\newcommand{\cm}{\textcolor{black}}

\newcommand{\beq}{\begin{equation}}
\newcommand{\eeq}{\end{equation}}
\newcommand{\bea}{\begin{eqnarray}} 
\newcommand{\eea}{\end{eqnarray}}
\newcommand{\ba}{\begin{array}}
\newcommand{\ea}{\end{array}}

\newcommand{\Msun}{M_{\odot}}

\newcommand{\mytextrm}[1]{{}}

\newcommand{\quarter}{\frac{1}{4}}
\newcommand{\half}{\frac{1}{2}}
\newcommand{\mat}[9] {
\left( 
\begin{array}{ccc}
#1 & #2 & #3 \\
#4 & #5 & #6 \\
#7 & #8 & #9
\end{array} \right)
}

\newcommand{\yml}{Y^m_l(\theta,\phi)}
\def\ltsima{$\; \buildrel < \over \sim \;$}
\def\simlt{\lower.5ex\hbox{\ltsima}}
\def\gtsima{$\; \buildrel > \over \sim \;$}
\def\simgt{\lower.5ex\hbox{\gtsima}}

\begin{document}
\title{Characterising gravitational wave stochastic background anisotropy with Pulsar Timing Arrays}


\author{C.~M.~F.~Mingarelli}
\affiliation{School of Physics and Astronomy, University of Birmingham, Edgbaston, Birmingham B15 2TT, UK}

\author{T. Sidery}
\affiliation{School of Physics and Astronomy, University of Birmingham, Edgbaston, Birmingham B15 2TT, UK}

\author{I.~Mandel}
\affiliation{School of Physics and Astronomy, University of Birmingham, Edgbaston, Birmingham B15 2TT, UK}

\author{A.~Vecchio}
\affiliation{School of Physics and Astronomy, University of Birmingham, Edgbaston, Birmingham B15 2TT, UK}

\date\today

\begin{abstract} 
Detecting a stochastic gravitational wave background, particularly radiation from individually unresolvable supermassive black hole binary systems, is one of the primary targets for Pulsar Timing Arrays. Increasingly more stringent upper limits are being set on these signals under the assumption that the background radiation is isotropic. However, some level of anisotropy may be present and the characterisation of the gravitational wave energy density
at different angular scales carries important information. We show that the standard analysis for isotropic backgrounds can be generalised in a conceptually straightforward way to the case of generic anisotropic background radiation by decomposing the angular distribution of the gravitational wave energy density
on the sky into multipole moments. We introduce the concept of generalised overlap reduction functions which characterise the effect of the anisotropy multipoles on the correlation of the timing residuals from the pulsars timed by a Pulsar Timing Array. In a search for a signal characterised by a generic anisotropy, the generalised overlap reduction functions play the role of the so-called Hellings and Downs curve used for isotropic radiation. We compute the generalised overlap reduction functions for a generic level of anisotropy and Pulsar Timing Array configuration. We also provide an order of magnitude estimate of the level of anisotropy that can be expected in the background generated by super-massive black hole binary systems.

\end{abstract}


\maketitle

\section{Introduction}
\label{s:intro}

The detection of gravitational waves (GWs) is one of the key scientific goals of Pulsar Timing Arrays (PTAs). A PTA uses a network of radio telescopes to regularly monitor stable millisecond pulsars, constituting a galactic-scale GW detector~\cite{PPTA, EPTA, IPTA, NANOGrav}. 
Gravitational radiation affects the propagation of radio pulses between a pulsar and a telescope at the Earth. The difference between the expected and actual time-of-arrival (TOA) of the pulses -- the so-called timing residuals -- carries information about the GWs~\cite{Sazhin:1978, Detweiler:1979, EstabrookWahlquist:1975}, which can be extracted by correlating the residuals from different pulsar pairs.
This type of GW detector is sensitive to radiation in the $10^{-9}-10^{-7}$~Hz frequency band, a portion of the spectrum in which a promising class of sources are super-massive black hole binary (SMBHB) systems with masses in the range of $\sim 10^7-10^9\Msun$ during their slow, adiabatic in-spiral phase~\cite{RajagopalRomani:1995, WyitheLoeb:2003, JaffeBacker:2003, SesanaVecchioColacino:2008, SesanaVecchioVolonteri:2008, WenEtAl:2011, Sesana:2012}. Other forms of radiation could be observable by PTAs, such as cosmic strings~\cite{PshirkovTuntsov:2010, SanidasBattyeStappers:2012, KuroyanagiEtAl:2012} and/or a background produced by other speculative processes in the early universe, see \textit{e.g.}~\cite{Zhao:2011}.

A PTA can be thought of as an all-sky monitor that is sensitive to radiation from the whole cosmic population of SMBHBs radiating in the relevant frequency band. The overwhelming majority of sources are individually unresolvable, but the incoherent superposition of the very weak radiation from the many binaries in the population gives rise to a stochastic background\footnote{It would be more appropriate to call this radiation a \textit{foreground}, but to be consistent with the established terminology we will keep referring to it as a \textit{background}.} whose detection is within reach of current or planned PTAs~\cite{SesanaVecchioColacino:2008, Sesana:2012, SiemensEtAl:2013}. In addition, some of the binaries may be sufficiently luminous to stand out above the diffuse background level and could be individually observed~\cite{SesanaVecchio:2010, YardleyEtAl:2010}.  The search for GWs from a SMBHB background~\cite{HellingsDowns:1983, JenetEtAl:2006, vanHaasterenEtAl:2011, DemorestEtAl:2012}  and from individual resolvable sources~\cite{JenetEtAl:2004, YardleyEtAl:2010, LeeEtAl:2011, BabakSesana:2012, EllisJenetMcLaughlin:2012, EllisSiemensCreighton:2012} has recently catalysed the PTA GW search effort, and it is plausible that in the next 5 to 10 years GWs could indeed be detected. If not, stringent constraints can be placed on aspects of the assembly history of SMBHBs~\cite{VolonteriHaardtMadau:2003, KoushiappasZentner:2006, MalbonEtAl:2007, YooEtAl:2007}.

In all the searches carried out so far, it has been assumed that the stochastic background, regardless of its origin, is isotropic~\cite{HellingsDowns:1983, JenetEtAl:2006, vanHaasterenEtAl:2011, DemorestEtAl:2012}. This is well justified if the background is produced by some physical processes in the early universe or is largely dominated by high-redshift sources. Under the assumption of isotropy, the correlated output from the data from any two pulsars in the array depends only on the angular separation of the pulsars and is known as the Hellings and Downs curve 
 ~\cite{HellingsDowns:1983}. However, a PTA also carries information about the angular distribution of the GW power on the sky. It is therefore important to address how this information is encoded in the data, and the implications for analysis approaches.
 In fact, if evidence for a signal is found in the data, testing the assumption of isotropy could be one of the methods to confirm its cosmological origin. If, on the other hand, one expects some deviations from isotropy, which may be the case for the SMBHB background created by a finite population~\cite{RaviEtAl:2012, CornishSesana:2013}, it is useful to be able to extract constraints on the underlying physical population. 

In this paper we show how the correlated output from pulsar pairs in a PTA is related to the anisotropy of the signal, \textit{i.e.} the angular distribution of GW power on the sky, and how one can extract this information by measuring the multipole moments that characterise the anisotropy level, following an analogous approach to those applied to the case of ground-based~\cite{AllenOttewill:1997} and space-based~\cite{Cornish:2002} laser interferometric observations. By doing this, we generalise the Hellings and Downs curve to an arbitrary angular distribution on the sky. We also provide an estimate for the expected level of anisotropy for the background produced by an arbitrary population of sources, and in particular, the population of SMBHB systems. 

The paper is organised as follows. In Section \ref{s:signal}, we review the basic concepts of a GW stochastic background, and we estimate the expected level of anisotropy in a background produced by a population of SMBHB systems. We show that at low frequencies, where the PTA sensitivity is optimal and the number of sources that contribute to the background is very large, the expected level of anisotropy is small, and likely undetectable. However towards the high-frequency end of the sensitivity window, where the actual number of sources decreases sharply, the anisotropy level could be significant, increasing at smaller angular scales. In Section~\ref{s:timingresiduals} we show that the present analysis approaches for isotropic backgrounds can be generalised in a conceptually straightforward way to the case of anisotropic signals by decomposing the angular distribution of the GW power on the sky into multipole moments. We introduce the concept of \textit{generalised overlap reduction functions}, which replace the Hellings and Downs curve.
Each one of these characterises the effect of a given anisotropy multipole on the correlation of the timing residuals from a pulsar pair.
In Section~ \ref{s:genHDC} we derive expressions for the generalised overlap reduction functions for an arbitrary stochastic background angular distribution on the sky and PTA configuration. This is essential for future analyses of PTA data which include anisotropy as part of the model. Section~\ref{s:concl} contains 
our conclusions and suggestions for future work. 

For the rest of the paper we will consider geometric units, and therefore set $c = G = 1$, unless otherwise specified.

\section{Stochastic backgrounds}
\label{s:signal}

Let us consider a plane wave expansion for the metric perturbation $h_{ij}(t,\vec x)$ produced by a stochastic background:
\beq
h_{ij}(t,\vec x) = \sum_A \int_{-\infty}^\infty \!\!\!\!df\ \int_{S^2} \!\!\!
d\hat\Omega\ h_A(f,\hat \Omega)\ e^{i2\pi f(t-\hat \Omega\cdot\vec x)}\ 
e_{ij}^A(\hat \Omega)\ ,
\label{e:h_ab}
\eeq
where $f$ is the frequency of the GWs, the index $A = +\,,\times$ labels the two independent polarisations, the spatial indices are $ i,j=1,2,3 $, 
the integral is on the two-sphere $S^2$, and our sign convention for the Fourier transform $\tilde g(f)$ of a generic function $g(t)$ follows the GW literature convention
\beq
\tilde g(f) = \int_{-\infty}^{+\infty} dt \, g(t)\,  e^{-i2\pi f t}\,. 
\label{e:ft}
\eeq
The unit vector $\hat\Omega$ identifies the propagation direction of a single gravitational wave plane, that can be decomposed over the GW polarisation tensors $e_{ij}^A(\hat\Omega)$ and the two independent polarisation amplitudes,  $h_A(t, \hat \Omega)$ or equivalently $h_A(f,\hat \Omega)$~\cite{MTW, AllenRomano:1999}: 

\begin{subequations}
\begin{align}
h_{ij}(t,\hat\Omega) & = e_{ij}^+(\hat\Omega) h_+(t,\hat\Omega) + e_{ij}^{\times}(\hat\Omega)\, h_\times(t,\hat\Omega)\,,
\label{e:hab}
\\
h_{ij}(f,\hat\Omega) & = e_{ij}^+(\hat\Omega) h_+(f,\hat\Omega) + e_{ij}^{\times}(\hat\Omega)\, h_\times(f,\hat\Omega).
\label{e:hab_f}
\end{align}
\end{subequations}

The polarisation tensors $e_{ij}^{A}(\hat\Omega)$ are uniquely defined once one specifies the wave principal axes described by the unit vectors $\hat m$ and $\hat n$: 
\begin{subequations}
\begin{align}
e_{ij}^+(\hat\Omega) &=  \hat m_i \hat m_j - \hat n_i \hat n_j\,,
\label{e:e+}
\\
e_{ij}^{\times}(\hat\Omega) &= \hat m_i \hat n_j + \hat n_i \hat m_j\,.
\label{e:ex}
\end{align}
\end{subequations}

For a stationary, Gaussian and unpolarised background the polarisation amplitudes satisfy the following statistical properties: 
\beq
\langle h^*_A(f,\hat{\Omega})  h_{A'}(f',\hat{\Omega}') \rangle = \delta^2(\hat{\Omega},\hat{\Omega}') \delta_{AA'} \delta(f - f') H(f) P(\hat\Omega)\,,
\label{e:hstat}
\eeq
where $\langle \cdot \rangle$ is the expectation value and $ \delta^2(\hat{\Omega},\hat{\Omega}')= \delta(\cos\theta-\cos\theta')\delta(\phi-\phi')$ is the covariant Dirac delta function on the two-sphere \cite{Finn2009}. This condition implies that the radiation from different directions are statistically independent. Moreover, we have factorised the power spectrum such that $P(f,\hat{\Omega})=H(f)P(\hat{\Omega})$, where the function $H(f)$ describes the spectral content of the radiation, and $P(\hat{\Omega})$ describes the angular distribution on the sky.

The mass-energy density in GWs is~\cite{AllenRomano:1999}
\beq
\rho_\mathrm{gw} = \frac{1}{32\pi} \langle \dot{h}_{ij}(t,\vec{x}) \dot{h}^{ij}(t,\vec{x}) \rangle\,,
\label{e:rho_gw}
\eeq
and using Eqs.~(\ref{e:h_ab}) and~(\ref{e:hstat}) we have:
\beq
\langle \dot{h}_{ij}(t,\vec{x}) \dot{h}^{ij}(t,\vec{x}) \rangle = 32\pi^2 \int d\hat \Omega P(\hat{\Omega}) \int_0^\infty df f^2 H(f)\,.
\eeq
The GW energy density, Eq.~(\ref{e:rho_gw}) is related to the more frequently used density parameter $\Omega_\mathrm{gw}(f)$ by:
\beq
\Omega_\mathrm{gw}(f) \equiv \frac{1}{\rho_c} \frac{d\rho_\mathrm{gw}}{d\ln f}\, ,
\label{e:Omega_gw}
\eeq
where $d\rho_\mathrm{gw}$ is the GW energy density in the infinitesimal band from $f$ to $f+df$, $\rho_c = 3H_0^2 / 8\pi$ is the critical density at the present epoch, and $H_0$ is the value of the Hubble parameter today.
 
Using Eqs.~(\ref{e:h_ab}), (\ref{e:hstat}) and~(\ref{e:rho_gw}) we can rewrite Eq.~(\ref{e:Omega_gw}) as
\beq
\Omega_\mathrm{gw}(f) = 
\frac{8 \pi^2}{3 H_0^2} f^3 H(f)  
\int d\hat{\Omega}\, P(\hat{\Omega})\,,
\label{e:Omega_gw1}
\eeq
which shows that the energy-density content of the background is the result of contributions from all the directions $\hat \Omega$. Each direction on the sky need not contribute to the background in the same way, and the function $P(\hat\Omega)$ describes the angular dependence (the ``hot'' and ``cold'' spots). As in 
~\cite{AllenOttewill:1997}, we decompose the angular distribution function on the basis of the spherical harmonic functions,
\beq
P(\hat{\Omega}) \equiv \sum_{lm} c_l^m Y_l^m(\hat{\Omega})\,,
\label{e:P_omega}
\eeq
where the sum is over $0 \le l < +\infty$, and $|m| \le l$. The coefficients $c_l^m$ are the multipole moments of the radiation which characterise the angular distribution of the background. We adopt the convention that the monopole moment is normalised as
\beq
c_0^0 = \sqrt{4 \pi}\,,
\label{e:c00}
\eeq
which yields
\beq
\int d\hat{\Omega}\, P(\hat{\Omega}) = 4 \pi\,.
\eeq
Eq.~(\ref{e:Omega_gw1}) now becomes
\beq
\Omega_\mathrm{gw}(f) = \frac{32 \pi^3}{3 H_0^2} f^3 H(f) \,.
\eeq
The frequency spectrum of the background, whether from SMBHBs or other sources or processes in the early Universe, is described by the function $H(f)$. The angular distribution of the radiation is encoded in the values of the radiation multipole moments $c_l^m$, which become unknown parameters in the analysis. In Section 3 and 4 we will show how the $c_l^m$'s enter the likelihood function of PTA timing residuals, and how an arbitrary angular distribution affects the correlation of radiation at any two pulsars timed by an array. This provides a way of measuring the multipole moments. In the remainder of this Section we provide an estimate of the expected level of anisotropy in a background generated by the population of SMBHB systems. 

In order to gain some insight into this problem, let us consider an idealised situation, constructed as follows. Let us assume that the universe is populated by identical sources with number density $n$. If we want to estimate the level of anisotropy, we need to estimate the expected value of the energy density in GWs coming from sources in a solid angle $d\Omega$ centred on a direction $\hat\Omega$ and compare it to the energy density produced by sources in a cone centred on a different direction $\hat\Omega^\prime$. For this example we consider a Euclidean, static universe (or equivalently sufficiently nearby sources, such that we do not take into account effects of expansion and redshift).

In a conical volume $dV = D^2 dD d\Omega$ within the solid angle $d\Omega$ and at distance between $D$ and $D + dD$, the expected number of sources which contribute to the background is: 
\beq
dN = n D^2 dD d\Omega\,.
\label{e:dN}
\eeq
The actual number of sources is then governed by Poisson statistics, with mean $\mu = dN$ and variance $\sigma^2 = dN$.  If the volume $dV$ is sufficiently small that $dN \ll 1$, then the probability of finding one source is 
\beq
P(1) =  dN e^{-dN} \approx dN.
\label{e:P(1)}
\eeq
Since the probability of having more than one source within this volume is negligible, the probability of finding no sources is simply $1 - P(1) = 1-dN$. 

The expected total number of sources, $\mu_N$, present in the whole volume within a solid angle $d \Omega$ between the minimum and maximum distance, $D_m$ and $D_M$, respectively (to be discussed later), is given by the sum of the contributions from each slice in the cone. Similarly, the variance $\sigma_N^2$ is the sum of the variances from each conical slice. We therefore obtain 
\begin{subequations}
\bea
\mu_N = \sigma_N^2 & = & \int_{D_m}^{D_M} n D^2 dD d\Omega\,,
\\
& = & \left(n \frac{4\pi}{3} D_M^3\right)\,\left(\frac{d\Omega}{4\pi}\right)\, \left[ 1 - \left(\frac{D_m}{D_M}\right)^3\right]\,.
\label{e:musigma_N}
\eea
\end{subequations}

We now want to compute the expected contribution to the GW energy density per frequency interval and its variance. The GW energy density of each source scales as $1/D^2$. If we assume that all the sources are identical -- the generalisation to a distribution of masses is straightforward, but is not needed to explain the key points -- we can write (with slight abuse of notation) the contribution to the energy density per source simply as
\beq
\frac{d\rho_\mathrm{gw}}{dN} = \frac{A}{D^2}\,,
\label{e:drhodN_toy}
\eeq
where $A$ is an appropriate constant factor, equal for all sources. 

The expected GW energy density from sources in a small conical volume $dV$ at distance $D$, again chosen so that it has a vanishingly small probability of having more than one source, $dN \ll 1$, see Eqs.~(\ref{e:dN}) and~(\ref{e:P(1)}), is 
\beq
\label{e:muD}
d\mu_\mathrm{gw}(D) \approx P(1) \frac{d\rho_\mathrm{gw}}{dN} \approx dN \frac{A}{D^2} = n A {dD} d\Omega\,,
\eeq
The variance of the energy density from sources in this conical volume is 
\beq
\label{e:sigmaD}
d\sigma^2_\mathrm{gw}(D) \approx P(1) \left( \frac{d\rho_\mathrm{gw}}{dN} \right)^2 - \left(\cm{d\mu_\mathrm{gw}}(D)\right)^2 \approx \frac{n A^2}{D^2}  {dD} d\Omega,
\eeq
where the last equality relies on the consistent application of the condition $dN \ll 1$ (which can always be satisfied by choosing a sufficiently small shell thickness $dD$).

We can now compute the expected contribution to the GW energy density $\mu_\mathrm{gw}$ and its variance $\sigma^2_\mathrm{gw}$
from all sources in a solid angle $d\Omega$.  The mean energy density and variance are given by the sum of contributions from all slices of thickness $dD$; using Eqs.~(\ref{e:muD}) and~(\ref{e:sigmaD}), this yields:
\begin{subequations}
\bea
\mu_\mathrm{gw} & = & \int_{D_m}^{D_M} \frac{d\mu_\mathrm{gw}}(D){dD}\ dD\,,
\\
& = & n A d\Omega \int_{D_m}^{D_M} dD\,,
\\
& = & n A D_M \left[1 - \frac{D_m}{D_M} \right]  d\Omega\,,
\eea
\end{subequations}
and, using the fact that the variance of a sum is the sum of variances, 
\begin{subequations}
\bea
\sigma^2_\mathrm{gw} & = & \int_{D_m}^{D_M} \frac{d \sigma^2_\mathrm{gw}(D)}{dD}\ dD\,,
\\
& = & n A^2 d\Omega \int_{D_m}^{D_M} \frac{dD}{D^2}\,,
\\
& = & n A^2 \left[\frac{D_M - D_m}{D_M D_m}\right]  d\Omega\,.
\eea
\end{subequations}
We define the level of anisotropy as the ratio of the standard deviation in the GW power emanating from a given solid angle to the expected power from that angle:
\bea
\frac{\sigma_\mathrm{gw}}{\mu_\mathrm{gw}} & = & \left( n d\Omega\right)^{-1/2}
\left[(D_M - D_m) D_M D_m\right]^{-1/2} 
\nonumber\\
& = & \left(n D_M^3 d\Omega\right)^{-1/2}
\left[\left(1 - \frac{D_m}{D_M}\right) \frac{D_m}{D_M}\right]^{-1/2}.
\label{e:level_anis}
\eea

We can now return to the choice of the minimal and maximal distance, $D_m$ and $D_M$.  The maximal distance at which sources can be located is set by cosmology and the history of SMBH formation.
Meanwhile, the minimal distance of interest to us, $D_m$, corresponds to the maximal distance at which individual binaries can be resolved.  Individually resolvable binaries can be subtracted from the data, and are treated separately from the stochastic background.  An individual source can be efficiently searched for with matched filtering techniques~\cite{BabakSesana:2012, EllisJenetMcLaughlin:2012, EllisSiemensCreighton:2012, PetiteauEtAl:2012}. Therefore, we expect the power necessary to detect a single SMBH binary to be significantly less than the power necessary to measure a stochastic background.  Thus, in order for a stochastic background to be detectable {\it after all individual sources that are presumed to be detectable up to distance $D_m$ are removed}, the total power in the background must be significantly greater than the power in the weakest individually resolvable source:
\beq
4\pi n A D_M \left[1 - \frac{D_m}{D_M} \right]  \gg \frac{A}{D_m^2}.
 \label{e:Dm}
\eeq
Another way to interpret the preceding condition is to consider the idealised situation when the stochastic background provides the dominant noise source: optimal matched filtering would make it possible to individually resolve and subtract coalescing SMBH binaries with signal power far below the noise (background) levels.

We can recast the condition on the detectability of a stochastic background, Eq.~(\ref{e:Dm}), as
\beq
\left(n D_M^3\right) \left(\frac{D_m}{D_M}\right)^2 \left[ 1 - \left(\frac{D_m}{D_M}\right)\right] \gg 1\,.
\label{e:stoch}
\eeq
If we define $y \equiv {D_m}/{D_M}$, where $0 \le y \le 1$, this condition yields
\beq
\left(n D_M^3\right)\, y^2\, (1 - y) \gg 1 \, ,
\label{e:cond2}
\eeq
where $n D_M^3$ is the total number of sources in the universe, modulo a factor of order unity. We can now rewrite the level of anisotropy~(\ref{e:level_anis}) in the following form:
\begin{subequations}
\bea
\frac{\sigma_\mathrm{gw}}{\mu_\mathrm{gw}} & = & \left\{\left(\frac{1}{d\Omega}\right)\,\left[\frac{y}{\left(n D_M^3\right)(1 - y) y^2}\right]\right\}^{1/2}
\label{e:level_anis1}
\\
& = & \left[\left(\frac{4\pi}{d\Omega}\right) \frac{\alpha(y)}{N_0}\right]^{1/2} \, ,
\label{e:level_anis2}
\eea
\end{subequations}
where $N_0 = (4\pi/3) n D_M^3 (1 - y^3)$ is the total number of sources that contribute to the background and $\alpha(y) \equiv (y^2 + y + 1)/(3 y)$. Note that by virtue of condition~(\ref{e:cond2}), the second term in Eq.(\ref{e:level_anis1}) is always smaller than unity whenever the stochastic background is detectable, and is actually $\ll 1$. The level of anisotropy scales as $N_0^{-1/2}$, and increases by going to small angular scales $d\Omega$. However, there is an observational limit on the angular resolution of PTAs which will prevent very small angular scales from being probed.  Furthermore, at smaller angular scales, the signal will be progressively dominated by a smaller number of, possibly individually unresolvable, sources.  The number of sources in a cone of solid angle $d\Omega$ is
\begin{subequations}
\bea
\mu_N &=& \frac{n D_M^3 d\Omega}{3}\left[ 1 - \left(\frac{D_m}{D_M}\right)^3\right],\\
&=&  \frac{d\Omega}{3}\,\left(n D_M^3\right) (1 - y^3),
\label{e:cond1}
\\
&=&  \left(\frac{d\Omega}{4\pi}\right) N_0 .
\label{e:cond1}
\eea
\end{subequations}
When this quantity is larger but not {\it much} larger than unity, we expect to be in the middle ground between searches for individual sources and standard stochastic-background searches.  If this  occurs on resolvable angular scales where anisotropy is significant (cf.~Eq.~(\ref{e:level_anis1}) and Eq.~(\ref{e:level_anis3}) below), it will be interesting to check the efficiency of current search pipelines in this regime.

Using the results from \textit{e.g.}~\cite{SesanaVecchioColacino:2008} we can provide an order-of-magnitude estimate of the expected level of anisotropy that characterises the SMBHB background. From Figure 4 of Ref.~\cite{SesanaVecchioColacino:2008} we can see that the total number of sources that contribute in a frequency interval of width $T_\mathrm{obs}$, where $T_\mathrm{obs}$ is the observation time, can be approximated as:
\beq
N_0 \approx 5\times 10^{5} \left(\frac{f}{10^{-8}\,\mathrm{Hz}}\right)^{-11/3}\,\left(\frac{5\,\mathrm{yr}}{T_\mathrm{obs}}\right) \, ,
\label{e:N0}
\eeq 
where we used the fact that, during a SMBHB inspiral, the time the binary spends in a given frequency band scales as $dt/df \propto f^{-11/3}$.  Substituting Eq.~(\ref{e:N0}) into Eq.~(\ref{e:level_anis2}) and converting between the average angular scale $d\Omega$ and the multipole moment index $l$ using  $d\Omega = 4\pi/2l$, we obtain:
\bea
\frac{\sigma_\mathrm{gw}(f)}{\mu_\mathrm{gw}(f)} & \approx & 3\times 10^{-3} \left(\frac{f}{10^{-8}\,\mathrm{Hz}}\right)^{11/6} \!\left(\frac{5\,\mathrm{yr}}{T_\mathrm{obs}}\right)^{-1/2}\!\left(\frac{l}{2}\right)^{1/2}\! \!\alpha^{1/2} \, ,
\nonumber
\\
& \approx & 0.2 \left(\frac{f}{10^{-7}\,\mathrm{Hz}}\right)^{11/6} \!\left(\frac{5\,\mathrm{yr}}{T_\mathrm{obs}}\right)^{-1/2}\!\left(\frac{l}{2}\right)^{1/2}\! \!\alpha^{1/2} \, .
\label{e:level_anis3}
\eea

There will be few SMBHBs beyond redshift $\sim 5$, and individual sources are likely to be resolvable up to redshift $\sim 1$, so sources that contribute to the stochastic background are within redshift range $\approx 1$--$5$,  see \textit{e.g.}~\cite{SesanaVecchioColacino:2008, SesanaVecchioVolonteri:2008}.  Therefore, both $y$ and $\alpha$ will be factors of order unity.  We have confirmed this with a more careful calculation that takes cosmology and the redshifting of gravitational waves into account; however, we note that our simplified treatment relied on a constant density (rate) of coalescing SMBHBs in the Universe, and on a fixed amplitude at a given frequency for all sources, which corresponds to the assumption of a fixed source mass. 

As expected, the level of anisotropy at low frequencies and large angular scales is small. However, it can become non-negligible, at the tens of percent level, at frequencies $\sim 10^{-7}$ Hz.

\section{Effect of anisotropy on timing residuals}
\label{s:timingresiduals}
GWs affect the time of arrival at the telescope of radio pulses from ultra-stable pulsars. Consider a pulsar with frequency $\nu_0$ whose location in the sky is described by the unit vector $\hat{p}$. The pulsar is at a distance $L$ from the Earth. \av{The effect of a GW source in the direction $-\hat\Omega$ generating a metric perturbation $h_{ij}(t,\hat{\Omega})$ is to affect the actual frequency $\nu$ at which the radio pulses are received at a telescope, according to}
\bea
z(t,\hat\Omega) & \equiv & \frac{\nu(t) - \nu_0}{\nu_0}
\nonumber\\
& = & \frac{1}{2} \frac{\hat p^i\hat p^j}{1+\hat \Omega \cdot \hat p} \Delta h_{ij}(t,\hat{\Omega})\,,
\label{e:z}
\eea
where

\beq
\Delta h_{ij}(t,\hat\Omega) \equiv h_{ij}(t,\hat{\Omega}) - h_{ij}(t_\mathrm{p},\hat{\Omega}) 
\label{e:deltah}
\eeq
is the difference between the metric perturbation at the Earth $h_{ij}(t,\hat{\Omega})$, the so-called {\it Earth term}, with coordinates $(t,\vec{x})$, and at the pulsar $h_{ij}(t_\mathrm{p},\hat{\Omega})$, the so-called {\it pulsar term}, with coordinates $(t_\mathrm{p},\vec{x}_p)$.\footnote{Note that the equivalent expression in~Ref.~\cite{AnholmEtAl:2009}, Eq. (9), has a sign error, as acknowledged by the authors, see the discussion of Eq (29) in Ref.~\cite{BookFlanagan:2011}.} We consider a frame in which 
\begin{subequations}
\begin{align}
t_p & = t_e - L = t - L \quad \quad \vec{x}_p = L \hat p \,,
\\
t_e & = t \quad \quad \vec{x}_e = 0\,,
\end{align}
\end{subequations}
where the indices ``e'' and ``p'' refer to the Earth and the pulsar. 
In this frame we can therefore write Eq~\eqref{e:deltah} using Eq~\eqref{e:hab_f} 
\bea
\Delta h_{ij}(t,\hat \Omega) = &&
\sum_A \int_{-\infty}^\infty df e_{ij}^A(\hat \Omega) \ h_A(f,\hat \Omega)\
\!e^{i 2 \pi f t}\ \!\!
\nonumber\\
&& 
\times \left[1 - e^{-i 2 \pi f L (1+ \hat \Omega \cdot \hat p)}\right]\, . \nonumber\\
\label{e:deltah1}
\eea

\av{The fractional frequency shift produced by a stochastic background is simply given by integrating Eq.~(\ref{e:z}) over all directions. Using Eq~(\ref{e:deltah1}), we obtain:
\bea 
\label{eq:freqShift}
z(t)  &=& \int d\hat\Omega\, z(t,\hat\Omega)
\nonumber \\
& = & \sum_A \int_{-\infty}^\infty df \int_{S^2}d\hat\Omega F^A(\hat\Omega) h_A(f,\hat \Omega)\nonumber\\
&&\times e^{i 2 \pi f t}\ \left[1 -  e^{-i 2 \pi f L (1+ \hat \Omega \cdot \hat p)} \right],
\eea
where $F^A(\hat\Omega)$ are the antenna beam patterns for each polarisation $A$, defined as
\beq
F^A(\hat\Omega) =\left[\frac{1}{2} \frac{\hat p^i\hat p^j}{1+\hat \Omega \cdot \hat p} \ e_{ij}^A(\hat \Omega)\right].
\label{e:F+Fx}
\eeq
The quantity that is actually observed is the time-residual $r(t)$, which is simply the time integral of Eq.~(\ref{eq:freqShift}):
\beq
r(t) = \int^t dt' z(t')\,.
\label{e:r}
\eeq
}
The search for a stochastic background contribution in PTA data relies on looking for correlations induced by GWs in the residuals from different pulsars. These correlations in turn depend on the spectrum $H(f)$ of the radiation, cf. Eqs~(\ref{e:Omega_gw1}) and~(\ref{e:hstat}), and the antenna beam pattern convolved with the angular distribution $P(\hat\Omega)$ of the \av{GW energy density} 
in the sky, cf. Eq.~(\ref{e:P_omega}).

Regardless of whether the analysis is carried out in a frequentist framework, and therefore one considers a detection statistic, see \textit{e.g.} Ref~\cite{AnholmEtAl:2009}, or one builds a Bayesian analysis \textit{e.g.}~\cite{VH}, the key physical quantity that is exploited is the correlation of the timing residuals for every pair of pulsars timed by a PTA. In a frequentist analysis this enters into a suitable detection statistic, whereas in a Bayesian framework it enters the likelihood function. The expected value of the correlation between a residual from a pulsar, say $a$, at time $t_j$, with that from a different pulsar, say $b$, at time $t_k$ depends on terms of the form:
\begin{widetext}
\bea
\langle r_a^*(t_j) r_b(t_k) \rangle & = &
\left\langle \int^{t_j} dt' \int^{t_k} dt''\cm{ z_a^*(t') z_b(t'') }\right\rangle
\nonumber\\
& = & \left\langle \int^{t_j} \!\!\!\! dt' \!\!\int^{t_k} \!\!\!\! dt''\!\! \int_{-\infty}^{+\infty} \!\!\!\! df'\!\! \int_{-\infty}^{+\infty} \!\!\!\! df''\cm{\tilde z_a^*(f') \tilde z_b(f'')}\, e^{- i 2\pi (f't' - f''t'')}  \right\rangle  
\nonumber\\
& = & \int^{t_j} dt' \int^{t_k} dt'' \int_{-\infty}^{+\infty} df e^{- i 2\pi f(t' - t'')} H(f)\, ^{(ab)}\Gamma(f).
\label{e:corr}
\eea
\av{In analogy with Ref.~\cite{AllenRomano:1999}, we define the quantity in the previous equation that depends on the relative location of the pulsars in the PTA, and the angular distribution of the GW energy density as the \textit{overlap reduction function}:}
\beq\label{eq:HDsol}
^{(ab)}\Gamma(f) \equiv \int d\hat\Omega P(\hat \Omega) \kappa_{ab}(f,\hat\Omega) \left[\sum_A F_a^A(\hat\Omega) F_b^A(\hat\Omega)\right],
\eeq
where
\beq
\kappa_{ab}(f,\hat\Omega) \equiv \left[ 1 - e^{i 2 \pi f L_a (1+ \hat \Omega \cdot \hat p_a)} \right] \left[ 1 - e^{-i 2 \pi f L_b (1+ \hat \Omega \cdot \hat p_b)}\right].
\label{eq:kappa}
\eeq
\end{widetext}
In Eq.~(\ref{e:corr}) $H(f)$ contains the information of the spectrum of radiation, and ${}^{(ab)}\Gamma(f)$ contains information about the angular distribution of GW background power. Under the assumption that the background is isotropic, ${}^{(ab)}\Gamma(f)$ is a known function that simply depends on the location of the pulsars timed by the array \cm{since  $P(\hat\Omega)$ is constant. In this case, the overlap reduction function~(\ref{eq:HDsol}) becomes:}
\beq
^{(ab)}\Gamma(f) = \int d\hat\Omega\, \kappa_{ab}(f,\hat\Omega)\, \sum_A F_a^A(\hat\Omega) F_b^A(\hat\Omega)\,,
\eeq
which is the result derived by Hellings and Downs in Ref.~\cite{HellingsDowns:1983} and is known (up to a normalisation constant) as the Hellings and Downs curve.

For an anisotropic background, whose angular power spectrum is unknown, $P(\hat\Omega)$ is a function of the unknown angular power distribution on the sky. We can generalise the concept of the overlap reduction function by decomposing $P(\hat \Omega)$ on the basis of spherical harmonic functions according to Eq.~(\ref{e:P_omega}). The weight of each of the components is given by an unknown coefficient $c_l^m$, which needs to be determined by the analysis. The overlap reduction function~(\ref{eq:HDsol}) becomes therefore:
\beq
^{(ab)}\Gamma(f) = \sum_{lm} c_l^m\, ^{(ab)}\Gamma_l^m(f)
\label{eq:HDdec}
\eeq
where 
\beq
^{(ab)}\Gamma_l^m(f) \equiv \int d\hat\Omega Y_l^m(\hat \Omega) \kappa_{ab}(f,\hat\Omega) \left[\sum_A F_a^A(\hat\Omega) F_b^A(\hat\Omega)\right]
\label{e:abGammalm}
\eeq
are the \textit{(complex-form) generalised overlap reduction functions}. Given an array of pulsars on the sky, the functions $^{(ab)}\Gamma_l^m$ are uniquely defined and known.

The generalisation of \textit{e.g.} the standard Bayesian analysis for an isotropic stochastic background such as the one reported in Ref.~\cite{VH} to the case in which the assumption of isotropy is relaxed is, at least conceptually, straightforward. The model parameters that describe the stochastic background are not only those that enter the frequency spectrum $H(f)$ -- for example the overall level and spectral index in the common case of a power-law parametrisation of $H(f)$, appropriate for the background from SMBHBs -- but also the coefficients that describe the angular distribution on the sky, that is, how much power is associated to each spherical harmonic decomposition of the overall signal. An initial implementation of this analysis is reported in Ref.~\cite{TaylorGair:anisotropy}.

Before we compute the expressions for the generalised overlap reduction functions, it is important to consider the function $\kappa_{ab}(f,\hat\Omega)$, defined in Eq.~(\ref{eq:kappa}) and present in Eqs.~(\ref{eq:HDsol}) and~(\ref{e:abGammalm}), which introduces the frequency dependence of the overlap reduction functions. From a physical point of view $\kappa_{ab}(f,\hat\Omega)$ encodes the fact that the correlation of the timing residuals carries information about both the Earth and pulsar terms for the two pulsars whose timing residuals are correlated. The relevant scale in the function $\kappa_{ab}(f,\hat\Omega)$ is
\beq
2 \pi f L (1+ \hat \Omega \cdot \hat p) = \cm{6.5\times 10^3 \left(\frac{f}{10^{-8}\,\mathrm{Hz}}\right)\!\! \left(\frac{L}{1\,\mathrm{kpc}}\right)\,
(1+ \hat \Omega \cdot \hat p)\,,}
\label{e:fL}
\eeq
which introduces rapid oscillations around unity~\cite{AnholmEtAl:2009} that depend on the distance and location to the pulsars. For all astrophysically relevant situations $fL \gg 1$, \av{see Eq.~(\ref{e:fL})}, and when one computes the integral in Eq.~(\ref{e:abGammalm}) the frequency dependent contributions to the integral rapidly average out to zero as the angle between the pulsar pairs, $\zeta$, increases. The generalised overlap reduction function Eq.~(\ref{e:abGammalm}) is therefore well approximated by 
\beq
^{(ab)}\Gamma_l^m \simeq (1 + \delta_{ab}) \int d\hat\Omega\, Y_l^m(\hat \Omega) \left[\sum_A F_a^A(\hat\Omega) F_b^A(\hat\Omega)\right] \,,
\label{e:abGammalm1}
\eeq
where $\delta_{ab}$ is the Kronecker delta. We will provide some more
details in Section~\ref{sec:pt_anisotropy}. Here we note that the
approximation~(\ref{e:abGammalm1}) is equivalent to considering only the
correlation of the Earth-term for two distinct pulsars. As we are considering many sources over the whole sky then
the pulsar terms will only contribute to the correlation if the distance between two pulsars
is of the order of one wavelength or less, and for the frequencies and pulsars
being considered this is only true for auto-correlation.  The auto-correlation term carries contributions from the Earth and pulsar terms, and therefore the value of of the integral is multiplied by a factor of 2. Note also, that the generalised overlap reduction function~(\ref{e:abGammalm1}) does not depend on frequency.

		The decompositions~(\ref{eq:HDdec}), (\ref{e:abGammalm}) and~(\ref{e:abGammalm1}) are based on the usual complex-basis spherical harmonic functions $Y_l^m(\hat \Omega)$, whose definitions are given in Section~\ref{sec:AniCorrFn}. One can alternatively consider a decomposition on a real basis $Y_{lm}(\hat \Omega)$, that are defined in terms of their complex analogs by\footnote{Here we adopt the convention that the real-form spherical harmonic functions and generalised overlap reduction functions  are written with indices $l$ and $m$ in the subscript, whereas in the complex-from, $m$ is raised as a superscript.}:

\beq
Y_{lm} = \left\{
\begin{array}{lr}
\frac{1}{\sqrt{2}} \left[Y_l^m + (-1)^m Y_l^{-m}\right]		&	\quad \quad 	m > 0\\
Y_l^0 										& 	\quad\quad 	m = 0 \\
\frac{1}{i\sqrt{2}} \left[Y_l^{-m} - (-1)^m Y_l^{m}\right] 	& 	\quad\quad 	m < 0\\
\end{array}
\right.
\eeq
Consequently, the real-form generalised overlap reduction functions are:
\beq
^{(ab)}\Gamma_{lm} =  \left\{
\begin{array}{lr}
\frac{1}{\sqrt{2}} \left[^{(ab)}\Gamma_l^m + (-1)^m \,^{(ab)}\Gamma_l^{-m}\right]		&	\quad \quad 	m > 0\\
^{(ab)}\Gamma_l^0 										& 	\quad\quad 	m = 0 \\
\frac{1}{i\sqrt{2}} \left[^{(ab)}\Gamma_l^{-m} - (-1)^m \,^{(ab)}\Gamma_l^{m}\right] 	& 	\quad\quad 	m < 0\\
\end{array}
\right.
\label{e:abGammalm_real}
\eeq
In the next Section we compute the $^{(ab)}\Gamma_l^m$'s for a generic pulsar pair and discuss their properties.

\section{Generalised overlap reduction functions}
\label{s:genHDC}

In this Section we compute the generalised overlap-reduction functions, Eq.~(\ref{e:abGammalm1}) for a generic pulsar pair and explore their properties. Anholm et al.~\cite{AnholmEtAl:2009} considered the particular case of the overlap-reduction function between two pulsars for radiation described by dipole anisotropy. Here we go beyond, and consider an arbitrary angular distribution of the background. Our approach is based on decomposing the power of the GW background at different angular scales onto spherical harmonics, cf. Eq.~(\ref{e:P_omega}) and for the specific case of a dipole distribution we show that our result is equivalent to the one presented in~\cite{AnholmEtAl:2009}.

\begin{figure}
	\centering
	\includegraphics[scale=0.47]{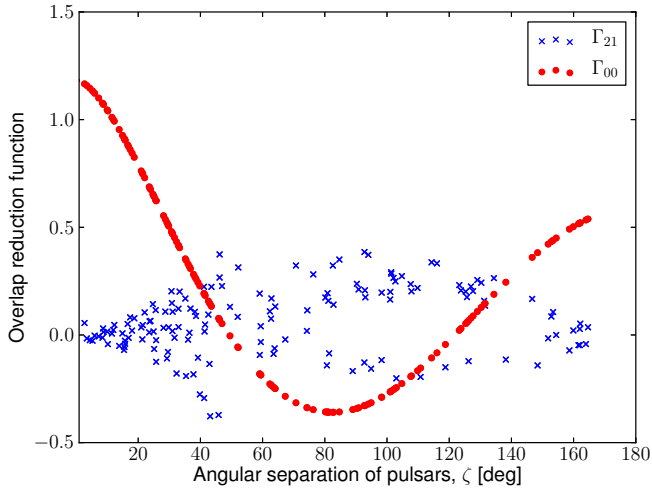}
	\caption{The real-value overlap reduction functions $\Gamma_{00}$ and $\Gamma_{21}$ for 18 EPTA pulsars in the cosmic rest-frame. Note that for illustrative purposes, we have not included the autocorrelation term ($\zeta=0$).} 
	\label{fig:EPTApulsars}
	\end{figure}

In the case of an isotropic background, pulsar pairs timed by a
PTA map uniquely into the Hellings and Downs curve. That is to say,
any pulsar pair is uniquely identified by an angular separation, which
in turn corresponds to a value of the overlap reduction
function. This is no longer the case for an anisotropic
distribution. For a given distribution of the GW power on the sky, the
generalised overlap reduction functions depend on the angular
separation between two pulsars \textit{and} their specific location in
the sky with respect to the background radiation. Equivalently, if one
considers two different pulsar pairs with the same angular separation
but different sky locations, the overlap reduction function that
describes the correlation between the two pulsars will be
different. To illustrate this, we show a selection of the best pulsars currently being timed by the European
  Pulsar Timing Array (EPTA) \cite{epta:web} \footnote{These are J0613$-$0200;
J1012$+$5307;
J1022$+$1001;
J1024$-$0719;
J1600$-$3053;
J1640$+$2224;
J1643$-$1224;
J1713$+$0747;
J1730$-$2304;
J1744$-$1134;
J1853$+$1303;
J1857$+$0943;
J1909$-$3744;
J1911$+$1347;
J1918$-$0642;
J1939$+$2134;
J2145$-$0750 and
J2317$+$1439.\\ These are the current  EPTA ``Priority 1" pulsars, however the prioritisation is subject to change.}   in Figure~\ref{fig:EPTApulsars},
  where we plot the real-valued overlap reduction functions, using Eq \eqref{e:abGammalm_real}, for the isotropic case
  and for $l = 2$ and $m = 1$. It can clearly be seen that the
  overlap reduction function no longer fits a single curve in the
  anisotropic case.

In our analysis we will closely follow the approach considered by Allen and Ottewill~\cite{AllenOttewill:1997}, who considered the equivalent problem in the case of ground-based laser interferometers.

\subsection{Choice of coordinate frames}

We introduce a ``cosmic rest-frame" where the angular dependency of the anisotropy is described, and a ``computational frame", in which some of the key expressions take a particularly simple form, and provide some intuitive clues into the problem. Given any two pulsars, say P$_a$ and P$_b$, we define the computational frame as the frame in which pulsar P$_a$ is on the z-axis, pulsar P$_b$ is in the $(x,z)$ plane, and their angular separation is denoted by $\zeta$. This is the standard frame that is used in \textit{e.g.}~\cite{AnholmEtAl:2009} to compute the Hellings and Downs curve for the isotropic case. Therefore, overlap reduction functions in the computational frame only depend on the pulsar pair's angular separation, $\zeta$. We now outline a method where one can rotate from the cosmic rest-frame to the computational frame, and vice versa, by means of rotation matrices.

Let us consider a generic vector ${\vec v}$, and let $v^{u}$ (unprimed) be the component in the cosmic rest-frame and $v^{u'}$ (primed) the component in the computational frame, which will be different for every pulsar pair. The components of the vector in the two different frames are related by:
\bea
v^{u'} & = & R_z(\gamma)\,R_y(\beta)\,R_z(\alpha) v^u,
\nonumber\\
& = & R(\alpha, \beta, \gamma)\,v^u,
\eea
where $R(\alpha, \beta, \gamma)$ is the rotation matrix given by:

\begin{align}
&R(\alpha, \beta, \gamma) =\\
&\mat{\cos \gamma}{ \sin \gamma}{0}{- \sin \gamma}{ \cos\gamma}{0}{0}{ 0 }{1 }\!\!\!\!
\mat{\cos\beta}{0}{- \sin\beta}{0}{1}{0}{\sin\beta}{0}{\cos\beta}\!\!\!\!
\mat{\cos \alpha}{ \sin \alpha}{0}{- \sin \alpha}{ \cos \alpha}{0}{0}{ 0 }{1 }.\nonumber
\end{align}
Indeed, we must carry out three rotations to go from the cosmic rest-frame to
the computational frame. If the pulsars P$_a$ and P$_b$ in the cosmic rest-frame have polar coordinates $(\theta_a,\phi_a$) and $(\theta_b,\phi_b$), respectively, the three angles of the rotations are:
\begin{subequations}
\begin{align}
\alpha & = \phi_a \,,
\\
\beta & = \theta_a\,,
\\
\tan \gamma & = \frac{\sin\theta_b \sin (\phi_b - \phi_a)}{\cos\theta_a \sin\theta_b \cos ( \phi_a - \phi_b)  - \sin\theta_a \cos\theta_b}\,.
\end{align}
\end{subequations}
The condition on $\gamma$ has two solutions within the range [0,2$\pi$] and we choose the one that gives a positive $x'$ coordinate in the computational frame for P$_b$.

Having calculated the relevant angles we can apply these to the rotation of spherical harmonics, where we know from Eq. (4.260) in Ref.~\cite{Arfken:1985}:
\begin{equation}
Y_{l}^m(\hat{\Omega}') = \sum_{k=-l}^{l}
D^{l}_{km}(\alpha,\beta,\gamma) Y_{l}^k(\hat{\Omega}),
\label{eq:Ylm_boring}
\end{equation}
and
\beq
Y_{l}^m(\hat\Omega) = \sum_{k = -l}^l \left[D^l_{mk}(\alpha, \beta, \gamma)\right]^* Y_{l}^k(\hat\Omega'),
\label{e:Ylm_prime}
\eeq
where equations~(\ref{eq:Ylm_boring}) and~(\ref{e:Ylm_prime}) rotate from the
computational frame into the cosmic rest-frame, and back to the computational frame, respectively.
The matrix $D^{l}_{mk}(\alpha,\beta,\gamma)$ is given by Eq. (4.12) in~\cite{Rose:1957}
\begin{equation}
D^{l}_{mk}(\alpha,\beta,\gamma) = e^{-im\alpha}d^{l}_{mk}(\beta) e^{-i k\gamma}\,,
\label{e:Dlmk}
\end{equation}
and  for $m\ge k$
\begin{widetext}
\begin{equation}
d^{l}_{mk}(\beta) = \left[ \frac{(l-k)!(l+m)!}{(l+k)!(l-m)!} \right]^{1/2} \frac{(\cos\frac{\beta}{2})^{2l+k-m} (-\sin \frac{\beta}{2})^{m-k}}{(m-k)!}
{}_{2} F_1 \left( m-l,-k-l;m-k+1;-\tan^2\frac{\beta}{2}\right)\,,
\end{equation}
\end{widetext}
where ${}_{2} F_1$ is the hypergeometric Gaussian function. For $m < k$, $d^l_{mk}$ can be derived from the unitary property, and yields 
\beq
d^l_{mk}(\beta) = d^l_{km}(-\beta) = (-1)^{m-k} d^l_{km}(\beta)\, ,
\eeq
as in Eq. (4.15) in Ref.~\cite{Rose:1957}. We also note that the $d^l_{mk}(\beta)$'s are real. Since ${}^{(ab)}\Gamma^m_l$ in Eq.~(\ref{e:abGammalm}) is a function of $Y^m_l$, we can now write the generalised overlap reduction function in the cosmic rest-frame as
\beq
^{(ab)}\Gamma_l^m(f) = \sum_{k=-l}^{l} [D^{l}_{mk}(\alpha,\beta,\gamma)]^* \Gamma'^k_l(f),
\label{e:abGammalm_rot}
\eeq
where ${}^{(ab)}\Gamma'^m_l(f)$ \av{(primed)} is the generalised overlap reduction function in the computational frame.

\subsection{Generalised overlap reduction functions in the computational frame} 
\label{sec:AniCorrFn}
In order to compute the generalised overlap reduction function in the cosmic rest-frame, Eq.~(\ref{e:abGammalm}) or~(\ref{e:abGammalm_real}), one needs to compute the relevant function in the computational frame then rotate it via Eq.~(\ref{e:abGammalm_rot}) using the matrix~(\ref{e:Dlmk}). Here we compute the generalised overlap reduction functions in the computational frame. For ease of notation, we drop the primes, but it understood that in this section all the analysis is done in the primed, computational frame.

The spherical harmonic function $\yml$ of order $m$ and degree $l$, $0\leq m\leq l$ is defined as
	\bea
	\yml&=&\sqrt{\frac{(2l+1)}{4\pi}\frac{(l-m)!}{(l+m)!}}P^m_l(\cos\theta)e^{im\phi}, \\
	&=&N^m_lP^m_l(\cos\theta)e^{im\phi}\label{eq:Oyml},
	\eea
where $0\leq\theta\leq\pi$ is the azimuthal angle and $0\leq\phi\leq 2 \pi$ is the polar angle and the $P^m_l(\cos\theta)$ are the associated Legendre polynomials 
\begin{subequations}
\label{e:P_aLegendre}
\bea
P^m_l(x)&=&\frac{(-1)^m}{2^l l!}(1-x^2)^{m/2}\frac{d^{l+m}}{dx^{l+m}}(x^2-1)^l\,,
\\
P^{-m}_l(x)&=&(-1)^m\frac{(l-m)!}{(l+m)!}P^m_l(x) \label{eq:minusM}\,,
\eea
\end{subequations}
and
	\beq
	 N^m_l=\sqrt{\frac{(2l+1)}{4\pi}\frac{(l-m)!}{(l\!+\!m)!}},
	 \label{e:normY}
	 \eeq
is the normalisation. The Hellings and Downs curve -- or equivalently the overlap reduction function for an isotropic background -- can be derived (up to a normalisation constant) setting $l = m = 0$, \textit{i.e.} $Y^0_0=1/\sqrt{4\pi}$. 

For each pair of pulsars, the computational frame is defined by the following geometry:
	\begin{subequations}
	\label{e:coord}
	\begin{align}
	\hat p_a		&=(0,0,1),\\
	\hat p_b		&=(\sin\zeta,0,\cos\zeta),\\
	\hat \Omega 	&=(\sin\theta\cos\phi,\sin\theta\sin\phi,\cos\theta),\\
	\hat m		&=(\sin\phi,-\cos\phi,0),\\
	\hat n		&=(\cos\theta\cos\phi,\cos\theta\sin\phi,-\sin\theta),
	\end{align}
	\end{subequations}
where $\zeta$ is the angular separation of the two pulsars, $\cos\zeta=\hat p_a \cdot \hat p_b$. In this frame $F_a^\times = 0$, and  
\av{Eq.~(\ref{e:abGammalm1})} reduces to 
	\beq
	{}^{(ab)}\Gamma^m_l=(1+\delta_{ab})\!\!\int_{S^2}d\hat\Omega\, Y_l^m(\hat\Omega) F^+_a(\hat\Omega)F^+_b(\hat\Omega).
        \label{eq:reductionSpheric}
	\eeq
With this choice of frame, the \cm{generalised} overlap reduction functions can be easily computed. It is worth pointing out that \textit{in this frame} the $\Gamma^m_l$'s are real $\forall l\,,m$, and therefore $\Gamma^{-m}_l=(-1)^m\Gamma^m_l$ since 
\av{$Y^{-m}_l=(-1)^m \left(Y^{m}_l\right)^*$}, where the star here denotes the complex conjugate. One then need only take into account the transformation properties of the associated Legendre polynomials defined in Eq.~(\ref{e:P_aLegendre}).

\begin{figure*}[ht!]
     \begin{center}
        \subfigure[Monopole $(l =0)$]{%
            \label{fig:HDcurve}
            \includegraphics[width=0.49\textwidth]{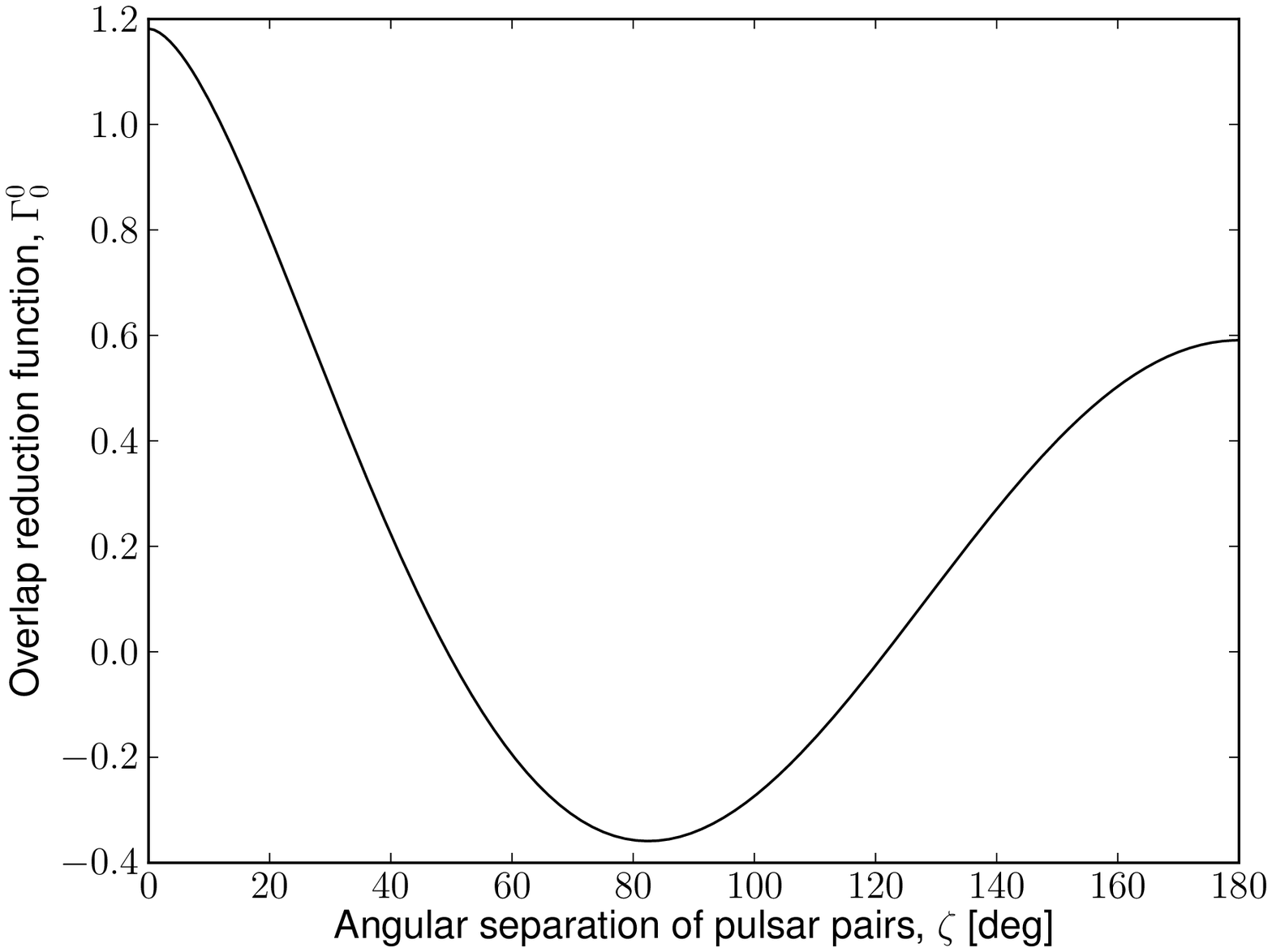}
        }%
        \subfigure[%
	Dipole $(l =1)$
        ]{%
           \label{fig:second}
           \includegraphics[width=0.49\textwidth]{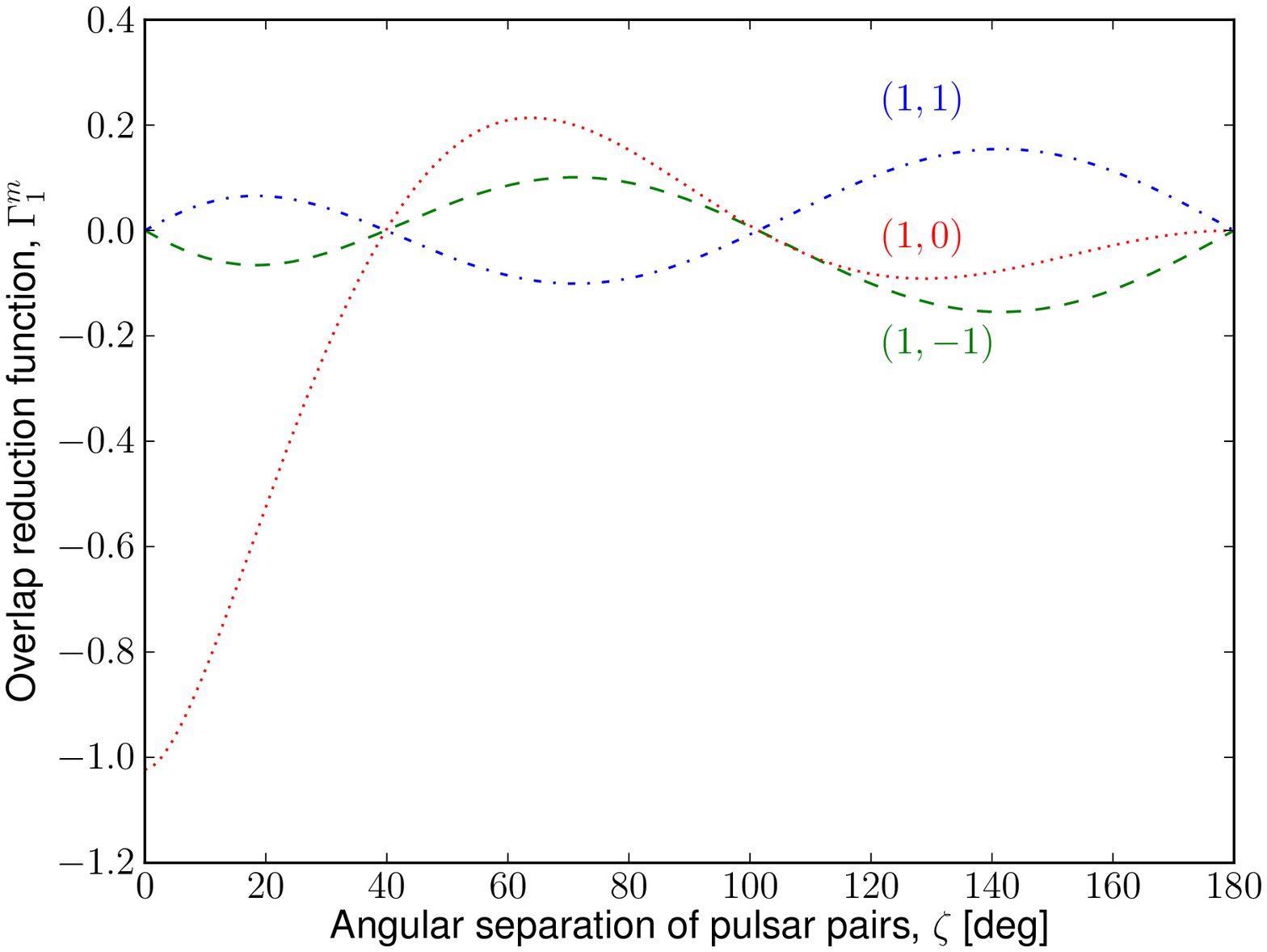}
        }\\ 
        \subfigure[%
        Quadrupole $(l =2)$]{%
            \label{fig:third}
            \includegraphics[width=0.49\textwidth]{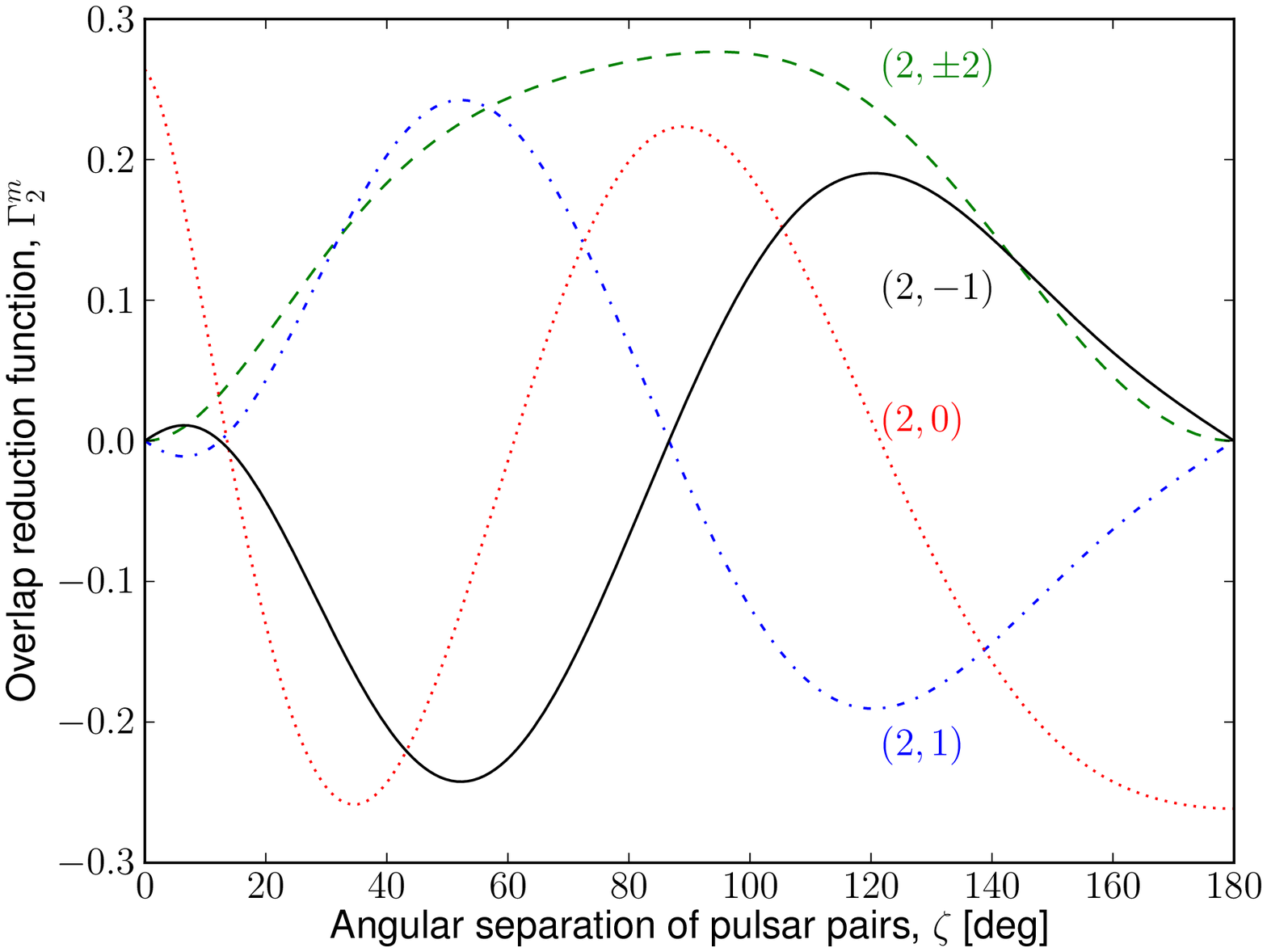}
        }%
        \subfigure[%
        Octupole $(l =3)$        ]{%
            \label{fig:fourth}
            \includegraphics[width=0.49\textwidth]{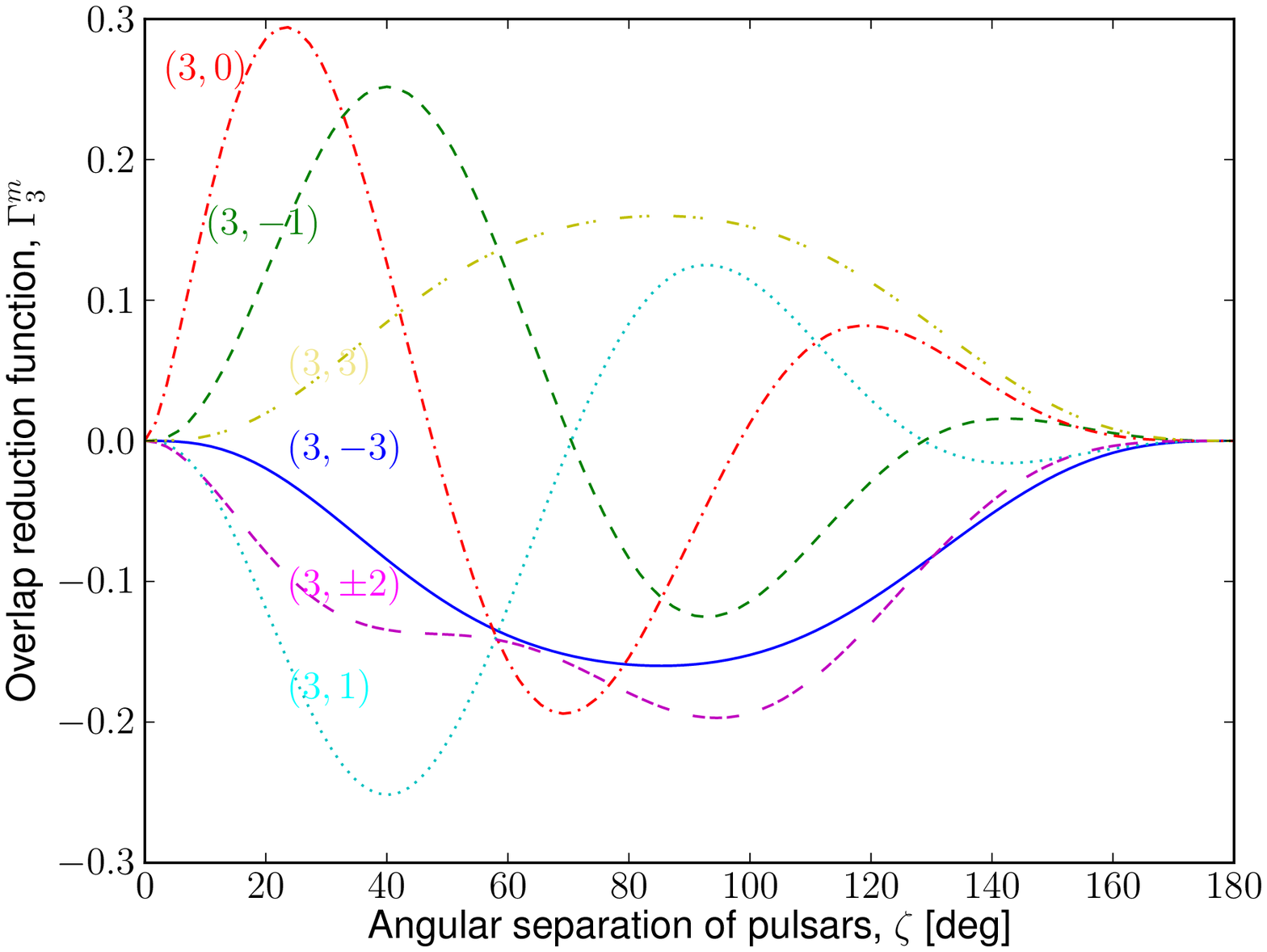}
        }%
    \end{center}
    \caption{%
        The Earth-term only, complex-valued, generalised overlap reduction functions $\Gamma_l^m$ in the computational frame for $l = 0\,, 1\,, 2\,, 3$ as a function of the angular separation of pulsar pairs. In the computational frame the functions are all real (see Appendix for more details) and $\Gamma^{-m}_l=(-1)^m\Gamma^m_l$. 
 For the case of the monopole ($l = 0$), the overlap reduction function $\Gamma_0^0$ is the Hellings and Downs curve up to the multiplicative constant $4\sqrt{\pi}/3$.
     }%
   \label{fig:subfigures}
\end{figure*}

\begin{figure*}[ht!]
     \begin{center}
        \subfigure[Difference of real parts] {%
		\label{fig:realfL10}
		\includegraphics[scale=0.4]{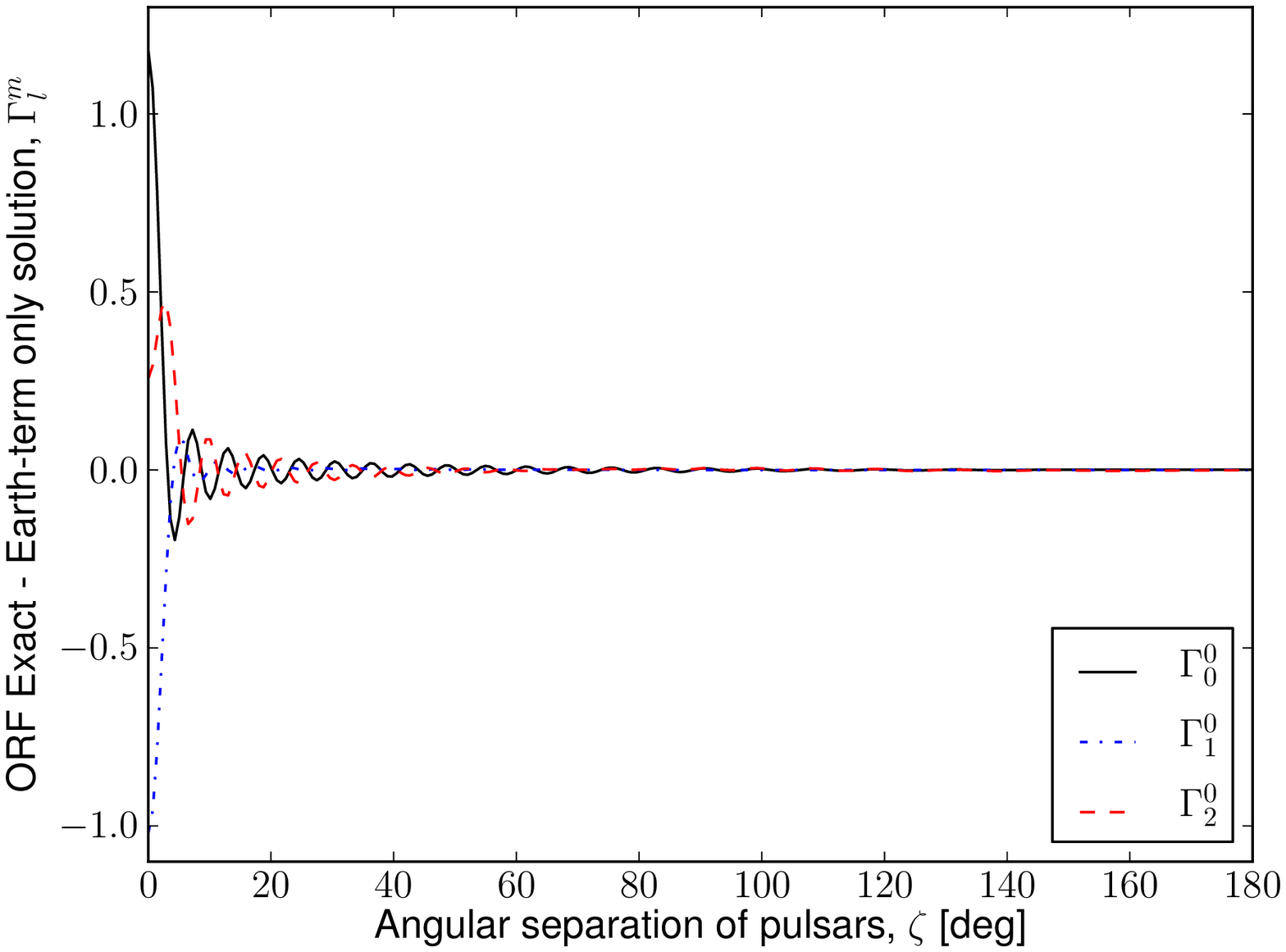}
        }%
	\subfigure[Imaginary part]{
		\label{fig:imfL10.eps}
	\includegraphics[scale=0.4]{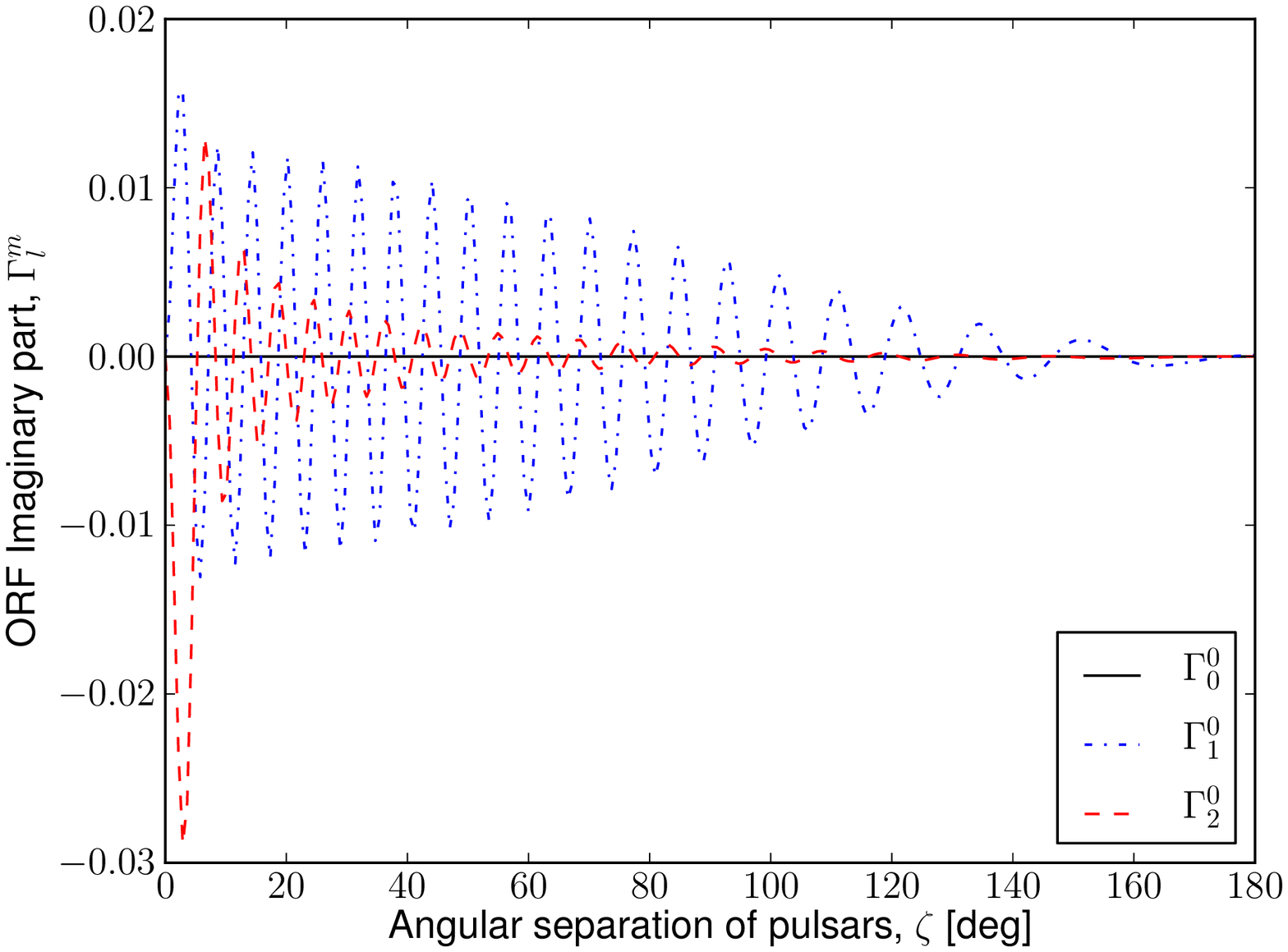}
	}
\end{center}

	\caption{Generalised overlap reduction functions (ORF) with the pulsar term. (a) The difference between the exact solution and the Earth term-only solution for $fL=10$ in the computational frame. These oscillations are already quite small for $\zeta=60^{\circ}$ and rapidly converge to zero for larger values of $\zeta$. (b) The value of the complex component of the pulsar term for $fL=10$ in the computational frame. Recall that the Earth-term only solution is always real, but introducing the pulsar term gives rise to complex-valued overlap reduction functions, even in the computational frame. Notice that these oscillations induced by the pulsar term are at least an order of magnitude smaller than the real part but do not, however, converge as quickly as the real component. The $\Gamma_0^0$ function has no imaginary component.}
\end{figure*}

In Appendix~\ref{sec:HDderivation} we provide comprehensive details of the derivations, whereas here we will just show the main results.
For the case $l=m=0$, Eq.~(\ref{eq:reductionSpheric}), we obtain the overlap reduction function for the case of an isotropic background (cf. Appendix~\ref{sec:HDsol}):
\beq \label{eq:HD}
{}^{(ab)}\Gamma^0_0=\frac{\sqrt{\pi}}{2}\left[1+\frac{\cos\zeta}{3}+4(1-\cos\zeta)\ln\left(\sin\frac{\zeta}{2}\right)\right](1+\delta_{ab}).
\eeq\
${}^{(ab)}\Gamma^0_0$ is the Hellings and Downs curve up to a multiplicative factor $4\sqrt{\pi}/3$. In fact the Hellings and Downs curve is normalised in such a way that is unity when one considers the auto-correlation of the timing residuals form the same pulsar ($a = b$ and therefore $\zeta = 0$). Note that for the isotropic case the rotation from the computational frame into the cosmic frame has no effect.

More generally, it is rather straightforward to compute analytical expressions for the case of a dipole ($l = 1$) anisotropy. In this case the generalised overlap reduction functions in the computational frame read (cf. Appendix~\ref{sec:dipAni}):
\begin{widetext}
\begin{subequations}
\label{e:gamma_dipole}
\begin{align}
{}^{(ab)}\Gamma^{-1}_1&=-\half\sqrt{\frac{\pi}{6}}\sin\zeta\left\{1+3 (1-\cos\zeta) \left[1+\frac{4}{(1+\cos\zeta)}\ln\left(\sin\frac{\zeta}{2}\right)\right] \right\}(1+\delta_{ab}),\\
{}^{(ab)}\Gamma^{0}_1&=-\half\sqrt{\frac{\pi}{3}}\left\{(1+\cos\zeta)+3 (1-\cos\zeta) \left[(1+\cos\zeta)+4\ln\left(\sin\frac{\zeta}{2}\right)\right]\right\}(1+\delta_{ab}),\\
{}^{(ab)}\Gamma^{1}_1&= - {}^{(ab)}\Gamma^{-1}_1\,,
\end{align}
\end{subequations}
\end{widetext}
and are shown in Figure~\ref{fig:second}. The generalised functions for $m = \pm 1$ satisfy $\Gamma^{-1}_1=-\Gamma^{1}_1$, since $m$ is odd. 
Eq.~(\ref{e:gamma_dipole}) are equivalent to the result obtained in~\cite{AnholmEtAl:2009}, where the dipole overlap reduction function is derived for a dipole in the direction:
\beq
\hat D = \left(\sin\alpha_a \cos\eta, \sin\alpha_a \sin\eta, \cos\alpha_a\right),
\eeq 
where
\beq
\hat D \cdot \hat p_a  =  \cos\alpha_a, \,\,\,\,\,\,\,\,  \hat D \cdot \hat p_b  =  \cos\alpha_b, 
\eeq
and so 
\beq\hat D \cdot \hat p_b =\cos\alpha_a \cos\zeta + \sin\alpha_a \sin \zeta \cos\eta.
\eeq
 In this case the function that describes the angular distribution in the sky is $P(\hat\Omega) = \hat D \cdot \hat\Omega$, therefore :
\bea
P(\hat\Omega)\!& = & \cos\alpha_a \cos\theta + \sin\alpha_a \sin\theta \cos(\phi - \eta).
\eea
Following our approach we can decompose $P(\hat\Omega)$ onto the basis of spherical harmonic functions and we obtain:
\begin{widetext}
\bea
P(\hat\Omega) & =  & 2\sqrt{\frac{\pi}{3}} \cos\alpha_a \!Y_1^0(\hat\Omega)
- \sqrt{\frac{2\pi}{3}} \!\left(\sin\alpha_a\! \cos\eta - i  \sin\alpha_a \!\sin\eta\right) Y_1^1(\hat\Omega)
+ \sqrt{\frac{2\pi}{3}}  \left(\sin\alpha_a \cos\eta + i  \sin\alpha_a \sin\eta\right) Y_1^{-1}(\hat\Omega) \nonumber\\
&= & \! 2\sqrt{\frac{\pi}{3}} \left\{\!\cos\alpha_a Y_{10}(\hat\Omega) 
\!-\! \sin\alpha_a\! \cos\eta Y_{11}(\hat\Omega)\!+\!\sin\alpha_a\! \sin\eta Y_{11}(\hat\Omega)\! \right\}\nonumber\\
\eea
\end{widetext}

The dipole overlap reduction function derived in~\cite{AnholmEtAl:2009} (see Eq. (C23) in Appendix 2), can therefore be written in terms of a linear combination of the generalised overlap reduction functions $^{ab}\Gamma_1^{-1}$, $^{ab}\Gamma_1^{0}$ and $^{ab}\Gamma_1^{1}$, or the analogous real expressions, and the actual values of the coefficients $c_1^{-1}$, $c_1^{0}$ and $c_1^{1}$ returned by the analysis provide the direction of the dipole moment that describes the radiation.

It is sufficiently straightforward to derive analytical expressions for the generalised overlap reduction function describing a quadrupole ($l = 2$) anisotropy (cf. Appendix~\ref{sec:quadAni}):

\begin{widetext}
\begin{subequations}
\label{e:gamma_quadrupole}
\begin{align}
&{}^{(ab)}\Gamma^{-2}_2=\Gamma^{2}_2,\nonumber\\
&{}^{(ab)}\Gamma^{-1}_2=-\Gamma^{1}_2,\nonumber\\
&{}^{(ab)}\Gamma^{0}_2=\frac{1}{3}\sqrt{\frac{\pi}{5}}\left\{\cos\zeta\!+\!\frac{15}{4} (1-\cos\zeta)\left[(1+\cos\zeta)(\cos\zeta\!+\!3)+\!8\ln\!\!\left(\sin\frac{\zeta}{2}\right)\right]\!\right\}(1+\delta_{ab}),\\
&{}^{(ab)}\Gamma^{1}_2=\quarter\sqrt{\frac{2\pi}{15}}\sin\zeta\!\left\{5\cos^2\zeta\!+\!15\cos\zeta\!-\!21\!-\!60\frac{(1-\cos\zeta)}{(1+\cos\zeta)}\ln\!\!\left(\sin\frac{\zeta}{2}\right)\right\}(1+\delta_{ab}),\\
&{}^{(ab)}\Gamma^{2}_2=-\quarter\sqrt{\frac{5\pi}{6}}\frac{(1-\cos\zeta)}{(1+\cos\zeta)}\left[(1+\cos\zeta)(\cos^2\zeta\!+\!4\cos\zeta-9)-24 (1-\cos\zeta)\ln\!\!\left(\sin\frac{\zeta}{2}\right) \right](1+\delta_{ab})\,,
\end{align}
\end{subequations}
\end{widetext}
which are shown in Figure~\ref{fig:third}. 
For higher order $l$ the integrals become sufficiently complex that we have not tried to derive analytical expressions. It is however easy to derive numerically the results, and an example for $l = 3$ is shown in Figure \ref{fig:fourth}.

\subsection{The pulsar term for generalised overlap reduction functions}
\label{sec:pt_anisotropy}

In our analysis we have approximated the generalised overlap reduction function, Eq.~(\ref{e:abGammalm}), as~(\ref{e:abGammalm1}) because current PTA analysis operates in the regime in which $fL \gg 1$. In other words, we have only considered the Earth-term contribution of the background in correlating data from different pulsars. At any given frequency, $\kappa_{ab}(f,\hat\Omega)$ introduces rapid oscillations that depend on the distance and location to the pulsars and the frequency of the gravitational radiation. When one integrates over the whole sky, all the possible directions of propagation of the background, the oscillations average to 1. Physically, this is a consequence of the fact that PTAs operate in the short-wavelength regime, that is the gravitational wavelength is much smaller than the distance to the pulsars.

In~\cite{AnholmEtAl:2009} it was shown that Eq.~(\ref{e:abGammalm1}) is an excellent approximation for $fL \gg 1$ for the isotropic (or monopole) case. The same is true for all the higher order moments $l$, and here we provide some examples. Let us consider $l =0, 1,2$ and the generalised overlap reduction functions which are non-zero at zero angular separation, that is $\Gamma^0_0$, $\Gamma^0_1$, and $\Gamma^0_2$. The functions which are zero at $\zeta=0$ have a very weak pulsar term dependence and are therefore not considered here. We will also make the assumption that the distance to both pulsars is the same.

The Earth term is always real for overlap reduction functions
calculated in the computational frame. By introducing the pulsar term,
the overlap reduction functions are in general complex; in fact, only
$\Gamma_l^0$ is real for all $l$. The pulsar term adds oscillations
which are at most twice the value of the Earth term for $\zeta=0$ and
are quickly attenuated as $\zeta$ increases. These oscillations can be seen in Figure \ref{fig:realfL10}, which shows the difference between the exact solutions of  Eq.~(\ref{e:abGammalm}) for $\Gamma^0_l$, where $l=0,1,2$, and the Earth-term only solutions Eq.~(\ref{e:abGammalm}), where we approximate $\kappa_{ab}\sim 1$. Note that these oscillations have almost converged to zero at $\zeta=60^{\circ}$ for $fL=10$. For larger values of $fL$, the pulsar term oscillations, such as the ones seen in Figure \ref{fig:realfL10}, become tighter and move to the left.

The imaginary part behaves in a similar oscillatory fashion. The oscillations in  Figure \ref{fig:imfL10.eps} are at least an order of magnitude smaller than those of the real part, and  can be thought of as a small change in phase. These oscillations converge much more slowly and in the case of $\Gamma^0_1$ they go to zero only at considerable angular separations.

\section{Conclusions}
\label{s:concl}

We have considered how an arbitrary level of anisotropy in the GW energy of a stochastic background affects the correlations of the data from pulsars in PTAs and the implications for analysis. In fact the characterisation of the GW power at different angular scales carries important information about the signal.

We have considered the relevant case of the
background from SMBHB systems. We have estimated that the
level of anisotropy is small, as one would expect, and likely
undetectable at present/near future sensitivity in the low-frequency
region, where PTAs have optimal sensitivity. The level of anisotropy
increases as one goes to higher frequencies, due to the fact that the
effective number of sources which dominate the signal
decreases. Anisotropy may therefore become important in a regime in
which the sources are still individually unresolvable (with the
exception of possibly a few), but the total number may not be sufficiently large to generate a smooth, diffuse background. This raises interesting questions regarding what is the optimal analysis strategy in this regime, which needs to be addressed. A detailed study of the anisotropy level that can be expected from astrophysically realistic populations of SMBHBs is currently in progress ~\cite{SesanaEtAl:anisotropy}. 

We have then shown that the present analysis techniques to search for isotropic stochastic backgrounds can be generalised to arbitrary levels of anisotropy by decomposing the angular distribution of the GW power on the sky into multipole moments. We have introduced the generalised overlap reduction functions $\Gamma_l^m$ that describe the correlation from the timing residuals from two pulsars for every $(l,m)$ anisotropy multipole. We have provided ready to use expressions for the $\Gamma_l^m$'s that can be used in the analysis of the data of the PTAs currently in operation and that are an essential element of an analysis pipeline aimed at this type of signal. A Bayesian analysis approach based on the formalism that we have presented here is being developed by Taylor and Gair~\cite{TaylorGair:anisotropy}. It is also important to note that some data analysis methods currently use ``compression" algorithms to speed up the processing of the data \cite{compression}. As a result of this, the high frequency sensitivity is compromised. This is the frequency band where anisotropy is more significant, and therefore future development of data analysis techniques will need to take this into account.

\section*{Acknowledgements}  
We would like to thank \cm{the referee for their thorough and careful report, which has undoubtedly made this work better. Furthermore, we would like to thank }S.~Taylor and J.~Gair for many invaluable discussions and suggestions. We would also like to thank A.~B.~Mingarelli, A.~Sesana and our colleagues of the European Pulsar Timing Array for comments. CMFM acknowledges the support of the Royal Astronomical Society. This research has been funded in part by an STFC grant.
\vspace{1cm}

\appendix

\section{Derivation of the generalised overlap reduction function}
\label{sec:HDderivation}
In this Appendix we provide details for the derivation of the analytical expressions of the generalised overlap reduction functions in the computational frame, \av{Eq~(\ref{eq:reductionSpheric})}, whose expressions are presented in Section~\ref{s:genHDC}. We begin by deriving identities and properties of integrals that will be used later in the derivations. We then derive the solutions for $l=0$ (isotropy) in Section \ref{sec:HDsol}, $l = 1$ (dipole anisotropy) in Section \ref{sec:dipAni}, and $l = 2$ (quadrupole anisotropy) in Section \ref{sec:quadAni}. 

We begin by choosing our reference frame as the ``computational frame" described in Section \ref{sec:AniCorrFn}. In this reference frame, the antenna beam patterns for pulsar $a$ and $b$ are:

	\begin{subequations}
	\label{e:F+Fx_comp}
	\begin{align}
	&F^\times_a=0,\\ 	
	& F^+_a=-\half(1-\cos\theta),\\
	&F^\times_b\!\!=\! \!\frac{(\sin\phi\,\sin\zeta\!)(\cos\theta\!\sin\zeta\!\cos\phi\!-\!\sin\theta\!\cos\zeta\!)}{1\!+\!\cos\theta\!\cos\zeta+\sin\theta\!\sin\zeta\cos\phi},\\
	&F^+_b\!\!=\!\!\half\frac{(\sin\phi\sin\zeta\!)^2\!\!-\!(\sin\zeta\!\cos\theta\!\cos\phi-\sin\theta\!\cos\zeta\!)^2}{\!1\!+\!\cos\theta\cos\zeta+\sin\theta\sin\zeta\cos\phi}.\nonumber\\
	\end{align}
	\end{subequations}
Substituting Eq.~(\ref{e:F+Fx_comp}) into Eq.~(\ref{eq:reductionSpheric}), the overlap reduction functions become:
\begin{widetext}
	\bea
	{}^{(ab)}\Gamma^m_l &=&-\quarter(1+\delta_{ab})\int_0^\pi \!\!\!d\theta\sin\theta\nonumber\\
	&\times&\!\!\!\int_0^{2\pi}\!\!\!\!d\phi
	Y^m_l\frac{(1-\cos\theta)(\sin^2\zeta\sin^2\phi-\sin^2\zeta\cos^2\theta\cos^2\phi-\cos^2\!\zeta\sin^2\theta+2\sin\zeta\!\cos\zeta\!\sin\theta\cos\theta\cos\phi )}
	{1+\sin\zeta\sin\theta\cos\phi+\cos\zeta\cos\theta}.
	\label{e:Gammalm_ex}
	\eea
\end{widetext}

One can write Eq \eqref{e:Gammalm_ex} as the sum of two integrals: 
\beq
{}^{(ab)}\Gamma^m_l=\quarter(Q^m_l+R^m_l)\label{eq:quarterGamma}(1+\delta_{ab})\,, 
\eeq 
where
\bea
&&Q^m_l=  N^m_l\int_0^\pi \!\!d\theta\sin\theta(1\!-\!\cos\theta)P^m_l(\cos\theta) \nonumber\\
&&\times \int_0^{2\pi}d\phi (1\!-\!\cos\zeta\!\cos\theta\!-\!\sin\zeta\!\sin\theta\cos\phi)e^{im\phi}
\label{eq:AppQ}
\eea
and
\bea 
	&&R^m_l\!=\!\!-N^m_l2\sin^2\zeta\!\!\int_0^{\pi}\!\!\!d\theta\sin\theta(1\!-\!\cos\theta)
				P^m_l(\cos\theta)I_m  \label{eq:appendixRLM}  \\
	&&I_m\equiv\!\!\int_0^{2\pi}\!\!\!d\phi\frac{e^{im\phi}\sin^2\phi}{1+\cos\zeta\!\cos\theta+\sin\zeta\!\sin\theta\cos\phi}\,,
	\label{eq:Im}
\eea
\av{and the constant $N_l^m$ is given by Eq.~(\ref{e:normY}).} 
The $Q^m_l$ portion of the overlap reduction function, Eq.~(\ref{eq:AppQ}), is only non-zero for $m=0,\pm 1$:
\beq
Q^m_l \ne 0 \quad \mathrm{iff}\, m=0,\pm 1\quad (\forall l)\,.
\label{e:lemma1}
\eeq
 This can be shown via integration by parts of the integral in $\phi$:
\bea
	&&\int_0^{2\pi}\!\!\!d\phi (1\!-\!\cos\zeta\!\cos\theta\!-\!\sin\zeta\!\sin\theta\cos\phi)e^{im\phi} = \nonumber\\
&&=-\int_0^{2\pi}d\phi \sin\zeta\!\sin\theta\cos\phi e^{im\phi}\\
	&&=-\sin\zeta\sin\theta\int_0^{2\pi}d\phi e^{im\phi} \cos\phi\\
&&=\sin\zeta\sin\theta\frac{im}{m^2-1}(e^{2i\pi m}-1)= 0 \quad (|m| \ge 2)\,.
\label{eq:zeroQ}
\eea
For $m = 0\,, \pm 1$, the integral in $\phi$ is handled as a special case: 
\bea
&& \int_0^{2\pi}\!\!\!d\phi (1\!-\!\cos\zeta\!\cos\theta\!-\!\sin\zeta\!\sin\theta\cos\phi)e^{im\phi} = 
\nonumber \\
&& = \left\{
\begin{array}{lc}
2\pi(1-\cos\zeta\cos\theta), 		& 	m = 0  		\label{eq:appMisZero}\\
-\pi\sin\zeta\sin\theta, 				& 	m = \pm 1 	\\
\end{array}
\right.
\eea
Note that the non-zero solutions given here are real-valued. We can now show that the generalised overlap reduction functions in the computational frame, given by Eq~(\ref{e:Gammalm_ex}) are real $\forall \,l\,,m$. 

We have just shown that the $Q^m_l$ are real, therefore it remains to prove that $R^m_l$, Eq.~\eqref{eq:appendixRLM}, is also real $\forall \,l\,,m$. The complex component is introduced via the $\phi$ dependence in Eq \eqref{eq:Im},
\bea
I_m&\equiv& \int_0^{2\pi}d\phi\frac{e^{im\phi}\sin^2\phi}{1+\cos\zeta\!\cos\theta+\sin\zeta\!\sin\theta\cos\phi},\\
&=&\int_0^{2\pi}d\phi\frac{\cos m\phi\sin^2\phi}{1+\cos\zeta\!\cos\theta+\sin\zeta\!\sin\theta\cos\phi}\nonumber\\
&+&i\int_0^{2\pi}d\phi\frac{\sin m\phi\sin^2\phi}{1+\cos\zeta\!\cos\theta+\sin\zeta\!\sin\theta\cos\phi}\,.
\eea
The final integral which is a function of $i$ can be written as an odd function over a symmetric interval for any value of $m$, hence it vanishes leaving only first, the real-valued, integral. Eq \eqref{eq:Im} can therefore be written as
\beq
I_m=\int_0^{2\pi}d\phi\frac{\cos m\phi\sin^2\phi}{1+\cos\zeta\!\cos\theta+\sin\zeta\!\sin\theta\cos\phi} \label{eq:realPhis},
\eeq
which is real-valued $\forall\, l\,,m$ in the computational reference frame.

Lastly we introduce an identity which helps one to readily solve a common integral involving Legendre polynomials. Formally, we show that for any $n$-times differentiable function $g(x)$ and Legendre polynomial $P_n(x)$, the following equality holds:
\beq
\int_{-1}^{1}dx\,g(x)P_n(x)=\frac{(-1)^n}{2^nn!}\int_{-1}^1 dx \, (x^2-1)^ng^{(n)}(x).
\label{e:l3}
\eeq
Using repeated applications of integration by parts, and using Rodrigues' formula for Legendre polynomials
\beq
P_n(x)=\frac{1}{2^nn!}D^n((x^2-1)^n)\,,
\eeq
where $D^n$ is the $n^{th}$ derivative with respect to $x$, the left-hand side of Eq~(\ref{e:l3}) can be written as:
\bea
	&&\int \, dx \,g(x)P_n(x)=g(x)\cdot\frac{1}{2^nn!}D^{n-1}((x^2-1)^n)\nonumber	\\
	&&-g'(x)\cdot\frac{1}{2^nn!}D^{n-2}((x^2-1)^n)+\label{eq:lemma}\cdots\nonumber	 \\ 
	&&+(-1)^{n-1}g^{(n-1)}(x)\cdot \frac{1}{2^nn!}D^{(n-n)}((x^2-1)^n)\nonumber	\\
	&&+\int dx\, (-1)^ng^n(x)\cdot\frac{1}{2^nn!}((x^2-1)^n).
\eea
We then evaluate Eq.~(\ref{eq:lemma}) over $[-1, 1]$ and note that in every boundary term, after the differentiations are performed, there is always a remaining term of the form $(x^2-1)^m$, for some $m$. Thus, this term vanishes at the end-points $ [-1, 1]$ leaving only the final integral term, thus proving Eq \eqref{e:l3}. We will make use of this identity regularly in the following sections describing dipole and quadrupole anisotropies.

\subsection{Isotropy}
\label{sec:HDsol}
We begin by solving the isotropic case which yields the Hellings and Downs curve, up to a normalisation constant. We therefore look to evaluate Eqs \eqref{eq:appendixRLM} and \eqref{eq:AppQ} with $l=m=0$:
\bea
Q^0_0&\!=\!&\frac{1}{\sqrt{4\pi}}\int_0^\pi \!\!d\theta  \sin\theta(1-\cos\theta)\nonumber \\
&&\times\!\int_0^{2\pi}\!\!\!d\phi(1-\cos\zeta\cos\theta-\sin\zeta\sin\theta\cos\phi),\\
&&=\frac{2\pi}{\sqrt{4\pi}}\int_0^\pi d\theta\sin\theta(1-\cos\theta)(1-\cos\theta\cos\zeta).\nonumber\\\label{eq:Q00}
\eea
Making the substitutions $x=\cos\theta$, $a'=\cos\zeta$ we can evaluate Eq.~(\ref{eq:Q00}) with the identity introduced in the preceding section using $g(x)=(1-x)(1-a'x)$:
	\bea
	Q^0_0&=&-\frac{2\pi}{\sqrt{4\pi}}\int_{+1}^{-1} dx(1-x)(1-a'x)P_0(x)\\
	&=&\sqrt{4\pi}\left(1+\frac{\cos\zeta}{3}\right).
	\eea
Next we solve for $R^0_0$
\beq\label{eq:ApR00}
R^0_0\!=\!-\frac{2}{\sqrt{4\pi}} \sin^2\zeta\!\int_0^\pi\!\!\! d\theta  \sin\theta(1-\cos\theta)I_{0},
\eeq
and using Eq \eqref{eq:realPhis} with $m=0$ we can write 
\beq
I_{0}\equiv\int_0^{2\pi}d\phi\frac{\sin^2\phi}{1+\cos\zeta\cos\theta+\sin\zeta\sin\theta\cos\phi}.
\eeq
This integral is evaluated via contour integration in \cite{AnholmEtAl:2009}, however we have used a symbolic program to evaluate it and have obtained the same result \footnote{Note that there is a sign typo in \cite{AnholmEtAl:2009}'s appendix in the equation above C9 (it does not have a number). Eq \eqref{eq:An1} has the correct sign. } \cm{
\bea
I_{0}&=&2\pi\frac{1+\cos\zeta\cos\theta-|\cos\zeta+\cos\theta|}{\sin^2\zeta\sin^2\theta}\label{eq:An1}\\
I_{0}&=&2\pi
 \left\{
\begin{array}{lc}
\left(\frac{1-\cos\zeta}{\sin^2\zeta}\right)\left(\frac{1-\cos\theta}{\sin^2\theta}\right), &		 	0<\theta<\pi-\zeta 	\\
\left(\frac{1+\cos\zeta}{\sin^2\zeta}\right)\left(\frac{1+\cos\theta}{\sin^2\theta}\right), &	 \pi-\zeta<\theta<\pi \label{eq:R00pm}\\
\end{array}
\right.
\eea 
}

We can now write down the final form of Eq \eqref{eq:ApR00}:
\bea
R^0_0&=&-\frac{4\pi(1-\cos\zeta)}{\sqrt{4\pi}}\!\int_0^{\pi-\zeta}\!\!\! d\theta  \frac{(1-\cos\theta)^2}{\sin\theta}\nonumber\\
&-&\frac{4\pi(1+\cos\zeta)}{\sqrt{4\pi}}\!\int_{\pi-\zeta}^\pi\!\!\! d\theta\sin\theta\\
&=&\sqrt{4\pi}(1-\cos\zeta)4\ln\left(\sin\frac{\zeta}{2}\right).
\eea
We simplify the final form by letting $\alpha=1+\cos\zeta$ and $\beta=1-\cos\zeta$, and will use these definitions extensively throughout the rest of this appendix.

 Using Eq \eqref{eq:quarterGamma}, one may write the isotropic solution to Eq \eqref{e:Gammalm_ex}:
\beq
{}^{(ab)}\Gamma^0_0=\frac{\sqrt{\pi}}{2}\left[1+\frac{\cos\zeta}{3}+4\beta\ln\left(\sin\frac{\zeta}{2}\right)\right](1+\delta_{ab})\,.
\eeq
This equation is the Hellings and Downs curve up to a multiplicative factor $4\sqrt{\pi}/3$. In fact the Hellings and Downs curve is normalised such that it is equal to 1 for  $\zeta = 0$, {\it i.e.} pulsar $a =$ pulsar $b$.

It is useful to note that when one sets $m=0$ and solves the above equations, one does so for any higher harmonic with $m=0$, as the integral in $\phi$ is solely a function of $m$. We can therefore write that for any $m=0$
 	\bea
	&&Q^0_l=2\pi N^0_l\int_0^\pi d\theta\sin\theta(1-\cos\theta)(1-\cos\zeta\cos\theta)P_l(\cos\theta)\label{eq:AQ0L},\nonumber\\
&&\\
	&&R^0_{l}=-4\pi N^0_l\beta\int_0^{\pi-\zeta}\!\!\!\!\!d\theta\frac{(1\!-\!\cos\theta)^2}{\sin\theta}P_l(\cos\theta)\nonumber\\
&&\;\;\;\;\;\;\;\;\;\;\;\;-4\pi N^0_l\alpha\int_{\pi-\zeta}^{\pi}\!\!\!\!\!\!d\theta\sin\theta\!P_l(\cos\theta). \label{eq:AR0L}
	\eea

\subsection{Dipole Anisotropy}
\label{sec:dipAni}
The dipole anisotropy is described by the $l=1$ and $m=0,\pm1$ spherical \av{harmonic functions.}
We therefore have non-zero solutions for all $Q^m_l$ and $R^m_l$. Here we derive the expressions for $\Gamma^0_1$ and $\Gamma^{\pm1}_1$. Beginning with $\Gamma^0_1$, one may easily compute $N^0_1=\sqrt{3/4\pi}$ and $P^0_1=\cos\theta$. Since $m=0$, the integral in $\phi$ is identical to that in the isotropic case for both $Q^0_1$ and $R^0_1$. We can also use~(\ref{e:l3}) to easily solve the integral in $\theta$, with $x=\cos\theta$ and $a'=\cos\zeta$:

\bea\label{eq:Q01}
Q^0_1&=&\sqrt{3\pi}\int_0^\pi d\theta\sin\theta(1-\cos\theta)(1-\cos\theta\cos\zeta)\cos\theta\nonumber\\
&=&\sqrt{3\pi}\int_{-1}^{+1}dx[a'x^2-x(a'+1)+1]x\\
&=&-2\sqrt{\frac{\pi}{3}}\alpha.
\eea
To evaluate $R^0_1$, we substitute $l=1$ into Eq.~\eqref{eq:AR0L}
\bea
R^0_1&=&-4\pi \sqrt{\frac{3}{4\pi}}\left[\beta\int_0^{\pi-\zeta}\!\!\!\!\!d\theta\frac{(1\!-\!\cos\theta)^2}{\sin\theta}\cos\theta\,\right.
\nonumber\\
& & \left. + \alpha\int_{\pi-\zeta}^{\pi}\!\!\!\!\!\!d\theta\sin\theta\cos\theta \right],\nonumber\\
&=&-2\sqrt{3\pi}\beta\left[\alpha+4\ln\left(\sin\frac{\zeta}{2}\right)\right],
\eea
so we can finally write
\beq
{}^{(ab)}\Gamma^0_1=-\half\sqrt{\frac{\pi}{3}}\left\{\alpha+3\beta\left[\alpha+4\ln\left(\sin\frac{\zeta}{2}\right)\right]\right\}(1+\delta_{ab}).
\eeq
To evaluate ${}^{(ab)}\Gamma^1_1$, we calculate $N^1_1=\sqrt{3/8\pi}$ and $P^1_1(\cos\theta)=-\sin\theta$ so that we can easily write
\bea
Q^1_1&=&\sqrt{\frac{3}{8\pi}}\int_0^\pi \!\!d\theta(-\sin^2\theta)(1\!-\!\cos\theta)\nonumber\\
&&\times \!\int_0^{2\pi}\!\!\!\!d\phi e^{i\phi}(1\!-\!\cos\zeta\!\cos\theta\!-\!\sin\zeta\!\sin\theta\cos\phi)\label{eq:appQ}\\
&=&\pi\sqrt{\frac{3}{8\pi}}\sin\zeta\int_0^\pi \!\!d\theta\sin^3\theta(1\!-\!\cos\theta)\\
&=&\sqrt{\frac{2\pi}{3}}\sin\zeta.\label{eq:AQ11}
\eea
Note that the solution of the integration in $\phi$ is valid for any $l$:
\beq
Q^1_l=-\pi N^1_l\sin\zeta \int_0^\pi d\theta \sin^2\theta(1-\cos\theta)P^1_l(\cos\theta)\label{eq:Q1allL}.
\eeq
We now turn our attention to $R^1_1$ and simplify the expression by substituting 
\begin{eqnarray*}
&&q=1+\cos\theta\cos\zeta, \\
&&r=\sin\theta\sin\zeta,
\end{eqnarray*}
\cm{noting that $\sqrt{q^2-r^2}=|\cos\theta+\cos\zeta|$}.  It follows that
\bea
	&&R^1_1=-2 \sqrt{\frac{3}{8\pi}}\sin^2\zeta\int_0^{\pi}\!\!\!d\theta\sin\theta(1\!-\!\cos\theta)(-\sin\theta)I_{1},\\
&&I_{1}\equiv\int_0^{2\pi}d\phi \frac{ \cos\phi\sin^2\phi}{q+r\cos\phi}.\\
&&=-\frac{\pi\left(-2q^3\!-\!r^2|\cos\theta\!+\!\cos\zeta|\!+\!2qr^2\!+\!2q^2|\cos\theta+\cos\zeta|\right)}{r^3|\cos\theta+\cos\zeta|}.\nonumber\\\label{eq:Aplusminus}
	\eea
As before, the value of Eq.~(\ref{eq:Aplusminus}) depends on where we are evaluating the integral in $\theta$: $\cos\theta+\cos\zeta$ is positive for $0\leq\theta\leq\pi-\zeta$ and negative for $\pi-\zeta\leq\theta\leq\pi$. We now factor Eq \eqref{eq:Aplusminus} \cm{considering $(\cos\theta+\cos\zeta)>0$}:

	\bea
	I_{1}&=&-\frac{\pi[q-(\cos\theta+\cos\zeta)]^2}{r^3},\\
	&=&-\frac{\pi}{\sin\!\theta\sin\!\zeta}\frac{(1\!-\!\cos\theta)(1\!-\!\cos\zeta)}{(1\!+\!\cos\theta)(1+\cos\zeta)}.
	\eea
\cm{The case where $(\cos\theta+\cos\zeta)<0$ is analogous. The complete expression for $I_{1}$ is therefore	
\bea
I_{1}&=&-\frac{\pi}{\sin\theta\sin\zeta}
 \left\{
\begin{array}{lc}
 \frac{(1-\cos\theta)(1-\cos\zeta)}{(1+\cos\theta)(1+\cos\zeta)},		& 	0<\theta<\pi-\zeta 	\\
\frac{(1+\cos\theta)(1+\cos\zeta)}{(1-\cos\theta)(1-\cos\zeta)}, 	& \pi-\zeta<\theta<\pi \label{eq:R00pm}\\
\end{array}
\right.
\eea
}	

Therefore, any $R^1_l$ can be written as:
	\begin{align} \label{eq:R1L}
	R^1_l&=+2\pi N^1_l \frac{\beta}{\alpha}\sin\zeta\int_0^{\pi-\zeta}d\theta\frac{(1-\cos\theta)^2}{1+\cos\theta}P^1_l(\cos\theta)\nonumber\\
	&+2\pi N^1_l \frac{\alpha}{\beta}\sin\zeta\int_{\pi-\zeta}^\pi d\theta(1+\cos\theta)P^1_l(\cos\theta)\Bigr.\,.
	\end{align}

For $m=1,l=1$, it is now straightforward to write
\bea
R^1_1&=&-\frac{\beta}{\alpha}\sqrt{\frac{3\pi}{2}}\sin\zeta \int_0^{\pi-\zeta}d\theta\frac{(1-\cos\theta)^2\sin\theta}{1+\cos\theta}\nonumber\\
&-&\frac{\alpha}{\beta} \sqrt{\frac{3\pi}{2}}\sin\zeta \int_{\pi-\zeta}^\pi d\theta \sin\theta(1+\cos\theta),\\
&=&2\beta\sqrt{\frac{3\pi}{2}}\sin\zeta\left[1+\frac{4}{\alpha}\ln\left(\sin\frac{\zeta}{2}\right)\right]\label{eq:AR11}.
\eea
Combining Eqs\eqref{eq:AQ11} and \eqref{eq:AR11} one finds the final expression for $\Gamma^1_1$:
\beq
{}^{(ab)}\Gamma^1_1=\half\sqrt{\frac{\pi}{6}}\sin\zeta\left\{1+3\beta\left[1+\frac{4}{\alpha}\ln\left(\sin\frac{\zeta}{2}\right)\right] \right\}(1+\delta_{ab}),
\eeq
and recall\av{ing} that ${}^{(ab)}\Gamma^{\av{-}m}_l={}^{(ab)}\Gamma^m_l(-1)^m$\av{, one obtains  ${}^{(ab)}\Gamma^{-1}_1= -{}^{(ab)}\Gamma^1_1$.}

\subsection{Quadrupole Anisotropy}
\label{sec:quadAni}
Quadrupole anisotropy is described in terms of the $l=2,m=0,\pm1,\pm2$ spherical \av{harmonic functions.} 
Two of these solutions are found immediately: since $l=2$, ${}^{(ab)}\Gamma^{-m}_2={}^{(ab)}\Gamma^m_2$. 
We now evaluate ${}^{(ab)}\Gamma^{|m|}_2$, beginning with ${}^{(ab)}\Gamma^0_2$, where $N^0_2=\sqrt{5/4\pi}$ and $P^0_2=1/2(3\cos^2\theta-1)$. Firstly we find $Q^0_2$ using \eqref{eq:AQ0L} with $l=2$
\bea
&&Q^0_2=\pi\sqrt{\frac{5}{4\pi}}\!\int_0^\pi \!\!\!\!d\theta\sin\theta(1\!-\!\cos\theta)(1\!-\!\cos\zeta\cos\theta)(3\cos^2\theta-1)\nonumber\\
&&=\frac{4}{3}\sqrt{\frac{\pi}{5}}\cos\zeta,
\eea
and $R^0_2$ can be found with \eqref{eq:AR0L} with $l=2$:
\bea
R^0_{2}&=&-2\pi \sqrt{\frac{5}{4\pi}}\beta\int_0^{\pi-\zeta}\!\!\!\!\!d\theta\frac{(1\!-\!\cos\theta)^2}{\sin\theta}(3\cos^2\theta-1)\!\nonumber\\ &-&2\pi \sqrt{\frac{5}{4\pi}}\alpha\int_{\pi-\zeta}^{\pi}\!\!\!\!\!\!d\theta\sin\theta(3\cos^2\theta-1),  \\
&=&\beta\sqrt{5\pi}\left[\cos^2\zeta+4\cos\zeta+3+8\ln\left(\sin\frac{\zeta}{2}\right)\right].
\eea
Combining these solutions we obtain:
\begin{widetext}
\beq
{}^{(ab)}\Gamma^{0}_2=\frac{1}{3}\sqrt{\frac{\pi}{5}}\left\{\cos\zeta\!+\!\frac{15\beta}{4}\left[\alpha(\cos\zeta\!+\!3)+\!8\ln\!\!\left(\sin\frac{\zeta}{2}\right)\right]\!\right\}(1+\delta_{ab}).
\eeq
\end{widetext}
Using analogous techniques, we can find an expression for  $\Gamma^1_2$. Here $N^1_2=\sqrt{5/24\pi}$ and $P^1_2=-3\cos\theta\sin\theta$, so $Q^1_2$ is given by substituting $l=2$ into Equation \eqref{eq:Q1allL}:
\bea
Q^1_2&=&3\pi\sqrt{\frac{5}{24\pi}}\sin\zeta\int_0^\pi \!\!\!\!d\theta \sin^3\theta\cos\theta(1-\cos\theta)\\
&=&-\sqrt{\frac{2\pi}{15}}\sin\zeta.
\eea
Equation \eqref{eq:R1L} is again used with $l=2$ to write $R^1_2$:
\begin{widetext}
\bea
R^1_2&=&-6\pi \sqrt{\frac{5}{24\pi}}\frac{\beta}{\alpha} \sin\zeta\int_0^{\pi-\zeta}d\theta\frac{(1-\cos\theta)^2\cos\theta\sin\theta}{1+\cos\theta}
-6\pi \sqrt{\frac{5}{24\pi}} \frac{\alpha}{\beta}\sin\zeta\int_{\pi-\zeta}^\pi d\theta(1+\cos\theta)\cos\theta\sin\theta\\
&=&-\frac{2\beta}{\alpha}\sqrt{\frac{5\pi}{6}}\sin\zeta\left[\alpha(\cos\zeta+4)+12\ln\left(\sin\frac{\zeta}{2}\right)\right].
\eea
Hence we write the final solution as:
\beq
{}^{(ab)}\Gamma^1_2=\quarter\sqrt{\frac{2\pi}{15}}\sin\zeta\!\left\{5\cos^2\zeta\!+\!15\cos\zeta\!-\!21\!-\!60\frac{\beta}{\alpha}\ln\!\!\left(\sin\frac{\zeta}{2}\right)\right\}(1+\delta_{ab}).
\eeq
\end{widetext}
Finally we write down the exact expression for ${}^{(ab)}\Gamma^2_2$. Recall that for $m=2$, $Q^2_2=0$ as shown in the introduction to this appendix. Here $N^2_2=\sqrt{5/96\pi}$ and $P^2_2=3\sin^2\theta$ and using $q$ and $r$ as previously defined we first write down the integral $I_2$:
\bea
&&I_{2}\equiv\int_0^{2\pi} d\phi \frac{\cos2\phi\sin^2\phi}{q+r\cos\phi},\\
&&=\frac{2\pi(\cos\theta+\cos\zeta)^2}{r^4|\cos\theta\!+\!\cos\zeta|}[2q|\cos\theta\!+\!\cos\zeta|\!-\!(\cos\theta\!+\!\cos\zeta)^2\!-\!q^2].\nonumber\\
\eea
This expression must be evaluated in 2 separate regimes, as before, where $\cos\theta+\cos\zeta$ is positive for $0\leq\theta\leq\pi-\zeta$ and negative for $\pi-\zeta\leq\theta\leq\pi$, i.e. 
\bea
 I_{2}&=&2\pi
 \left\{
\begin{array}{lc}
\frac{-(\cos\theta\!+\!\cos\zeta)}{(1+\cos\theta)^2(1+\cos\zeta)^2} ,		& 	0<\theta<\pi-\zeta 	\\
\frac{(\cos\theta\!+\!\cos\zeta)}{(1-\cos\theta)^2(1-\cos\zeta)^2} , 	& \pi-\zeta<\theta<\pi \\
\end{array}
\right.
\eea

Therefore:
\begin{widetext}
\bea
&&{}^{(ab)}\Gamma^2_2=\frac{3}{4}\sqrt{\frac{5\pi}{6}}\sin^2\zeta\int_0^{\pi-\zeta} d\theta \frac{\sin^3\theta(1-\cos\theta)(\cos\theta+\cos\zeta)}{\alpha^2(1+\cos\theta)^2}(1+\delta_{ab})-\frac{3}{4}\sqrt{\frac{5\pi}{6}}\sin^2\zeta\int_{\pi-\zeta}^\pi  d\theta \frac{\sin^3\theta(\cos\theta+\cos\zeta)}{\beta^2(1-\cos\theta)}(1+\delta_{ab}),\nonumber\\
&&{}^{(ab)}\Gamma^2_2=-\quarter\sqrt{\frac{5\pi}{6}}\frac{\beta}{\alpha}\left\{\alpha(\cos^2\zeta+4\cos\zeta-9)-24\beta\ln\left(\sin\frac{\zeta}{2}\right)\right\}(1+\delta_{ab}).
\eea
\end{widetext}

\end{document}